\documentclass[prd,twocolumn,showpacs,floatfix,superscriptaddress,nofootinbib]{revtex4}
\usepackage[utf8]{inputenc}
\usepackage{graphicx}
\usepackage{epsfig}
\usepackage{bm}
\usepackage{amsfonts}
\usepackage[T1]{fontenc}
\usepackage{amssymb}
\usepackage{float}
\usepackage{amsmath}
\usepackage{dcolumn}
\providecommand{\abs}[1]{\lvert#1\rvert}

\usepackage{cancel}
\usepackage[colorlinks]{hyperref}
\usepackage[usenames,dvipsnames]{color}
\hypersetup{
     breaklinks=true,
    pdfstartview={FitH},    
    colorlinks=true,       
    linkcolor=blue,          
    citecolor=red,        
    filecolor=magenta,      
    urlcolor=blue,           
    anchorcolor=green,      
    linktocpage=true
}

\def\doi{http://doi.org}

\def\be{\begin{equation*}}
\def\ee{\end{equation*}}

\begin{document}

\title{Study of Null Geodesics and their Stability in Horndeski Black  Holes}
\author{D. A. Carvajal}
\email{diego.carvajala@usach.cl} \affiliation{Instituto de F\'{i}sica, 
 Pontificia Universidad Cat\'{o}lica de Chile, Avenida Vicuña Mackenna 4860, Santiago, Chile.}
 \affiliation{Departamento de Matemática y Ciencia de la Computación, Universidad de Santiago de Chile, Las Sophoras 173, Santiago, Chile.}
 
\author{P. A. Gonz\'{a}lez}
\email{pablo.gonzalez@udp.cl} \affiliation{Facultad de
Ingenier\'{i}a y Ciencias, Universidad Diego Portales, Avenida Ej\'{e}rcito
Libertador 441, Casilla 298-V, Santiago, Chile.}
\author{Marco Olivares}
\email{marco.olivaresr@mail.udp.cl}
\affiliation{Facultad de
Ingenier\'{i}a y Ciencias, Universidad Diego Portales, Avenida Ej\'{e}rcito
Libertador 441, Casilla 298-V, Santiago, Chile.}
\author{Eleftherios Papantonopoulos}
\email{lpapa@central.ntua.gr}
\affiliation{Physics Division, School of Applied Mathematical and Physical Sciences, National Technical University of Athens, 15780 Zografou Campus,
    Athens, Greece.}
\author{Yerko V\'{a}squez}
\email{yvasquez@userena.cl}
\affiliation{Departamento de F\'{\i}sica, Facultad de Ciencias, Universidad de La Serena,\\
Avenida Cisternas 1200, La Serena, Chile.}
\date{\today}

\begin{abstract}

We study the motion of particles in the background of a scalar-tensor theory of gravity in which the scalar field is kinetically coupled to the Einstein tensor and we present the null geodesic structure for asymptotically flat, AdS, and dS Horndeski black holes, studying the effect of the cosmological constant on the orbits. Also, we consider three classical tests of gravity in the solar system, such as the bending of the light, the gravitational redshift, and the Shapiro time delay in order to constraint the coupling parameters of the scalar field to gravity.  Calculating the Lyapunov exponent we explore the stability of these geodesics for various values of the cosmological constant.  

\end{abstract}

\maketitle


\tableofcontents

\newpage


\section{Introduction}

Modified theories of gravity have been recently developed to address certain inconsistencies in General Relativity (GR) and to explain observations related to dark matter and dark energy. These theories propose alterations to the GR at both short and large distances, with the aim of offering a viable gravitational framework. Recent detections of gravitational waves (GWs) \cite{Abbott:2016blz,Abbott:2016nmj,Abbott:2017vtc,Abbott:2017oio,TheLIGOScientific:2017qsa} have opened new avenues for testing alternative gravity theories and distinguishing them from GR. Therefore, it is crucial to examine the compact objects predicted by various modified gravity theories and the potential GW signatures they may produce. Additionally, investigating classical solar system tests, such as light deflection, planetary perihelion shifts, and gravitational time delays, can reveal discrepancies between GR and observational data.

Scalar-tensor theories \cite{Fujii:2003pa}, among the simplest and most extensively studied modifications of GR, introduce a scalar field coupled to gravity. This coupling leads to the formation of black holes and compact objects dressed with a hairy matter distribution. Horndeski's theory \cite{Horndeski:1974wa}, a prominent scalar-tensor framework, is distinguished by second-order field equations that prevent ghost instabilities \cite{Ostrogradsky:1850fid} and preserve classical Galilean symmetry \cite{Nicolis:2008in,Deffayet:2009wt}. The theory has been analyzed across both short and cosmological distances. At shorter scales, subclasses featuring a scalar field kinetically coupled to the Einstein tensor admit local black hole solutions \cite{Kolyvaris:2011fk,Rinaldi:2012vy,Kolyvaris:2013zfa,Babichev:2013cya,Charmousis:2014zaa}. Moreover, other subclasses of the theory also admit black hole solutions \cite{Babichev:2017guv, Bergliaffa:2021diw, Walia:2021emv}. A study of the superradiance phenomenon of a rotating Horndeski black hole was performed in \cite{Jha:2022tdl}. On cosmological scales, the derivative coupling in Horndeski theories acts as a friction term during the inflationary period of the universe \cite{Amendola:1993uh,Sushkov:2009hk,Germani:2010hd,Saridakis:2010mf,Huang:2014awa,Yang:2015pga,Koutsoumbas:2013boa}. This coupling introduces a mass scale that, at large distances, can be constrained by GW observations.  Moreover, large black-hole scalar charges induced by cosmology in Horndeski theories was studied in Ref. \cite{Babichev:2025ric}, and extended thermodynamic analysis of a charged Horndeski black hole in Ref. \cite{Myung:2025wmw}.

In models in which dark energy is represented by a scalar field coupled to the Einstein tensor, studies have shown that the propagation speed of GWs differs from the speed of light \cite{Germani:2010gm,Germani:2011ua}. This discrepancy allows for constraints on the derivative coupling parameter and serves as a test for the applicability of Horndeski theories at cosmological scales \cite{Lombriser:2015sxa,Lombriser:2016yzn,Bettoni:2016mij,Baker:2017hug,Creminelli:2017sry,Sakstein:2017xjx,Ezquiaga:2017ekz}.

The measurement of the speed of GWs by GW170817 and GRB170817A provided an upper bound on the speed of GWs $c_{gw}/c-1\le 7\times 10^{-16}$ \cite{Monitor:2017mdv}. Assuming that the peak of the GW signal and the gamma-ray burst (GRB170817A) were emitted simultaneously, a lower bound can be derived $c_{gw}/c-1>-3\times 10^{-15}$ \cite{Monitor:2017mdv}, from which we can safely conclude that $c_{gw}=c$. Precise measurements of the propagation speed of GWs are a powerful tool for constraining the applicability of Horndeski theory. Specifically, in Horndeski theory \cite{Horndeski:1974wa} and its generalizations \cite{Deffayet:2009mn}, the functions of the scalar field $\phi$ and its kinetic energy $X = -\frac{1}{2} \partial_\mu \phi \partial^\mu \phi$, $G_4(\phi, X)$ and $G_5(\phi, X)$ should must be constrained to remain consistent with the aforementioned observations. This is because these terms provide the kinetic energy of the scalar field coupled to gravity, and they influence the speed of GWs. The term $G_5(\phi, X)$ represents the general coupling of the scalar field to the Einstein tensor, and in \cite{Gong:2017kim}, assuming that the scalar field plays the role of dark energy, a lower bound on the mass scale present in this term was found. Combining the constraints from inflation, the energy scale of the derivative coupling is bounded to be $10^{15}\text{GeV}\gg M \gtrsim 2\times 10^{-35}$ GeV.

Modified gravity theories can also be contrasted with the predictions of GR at relatively small scales. Observations within the solar system, such as light deflection, the perihelion shift of planets, and gravitational time delay, are well-explained by GR. To investigate these phenomena, one needs to compute the geodesics for particle motion around a black hole background. In \cite{Chakraborty:2012sd}, the perihelion precession of the planetary orbits and the bending angle of the null geodesics were calculated for various gravity theories in string-inspired models. The effects of gravity on the solar system have also been studied in black hole AdS geometries by analyzing the motion of particles in AdS spacetime \cite{Cruz:2004ts, Vasudevan:2005js, Hackmann:2008zz, Hackmann:2008zza, Olivares:2011xb, Cruz:2011yr, Larranaga:2011fp, Villanueva:2013zta}. The motion of massless and massive particles in the background of four-dimensional asymptotically AdS black holes with scalar hair was explored in \cite{Gonzalez:2013aca} and \cite{Gonzalez:2015jna}, see \cite{Ahmed:2025sav, Ahmed:2025evv, Ahmed:2025iqz} for other asymptotically AdS spacetime. The geodesics were computed numerically, and the differences in the particle dynamics between the black holes with scalar hair and their no-hair limit were discussed. In the context of both solar system and astrophysical scenarios, spherically symmetric solutions arising from the coupling of the Gauss-Bonnet term with a scalar field were analyzed in \cite{Bhattacharya:2016naa}. The motion of particles in the background of a scalar-tensor theory of gravity in which the scalar field is kinetically coupled to the Einstein tensor was studied in \cite{Gonzalez:2020vzl}, and the value of the derivative parameter was constrained by solar system tests. In addition, the geodesic motion in Euclidean Schwarzschild geometry was investigated in \cite{Battista:2022krl}.

The study of the motion of charged or uncharged particles outside the horizon of a black hole (BH) can give us information about whether these particles follow stable or unstable circular orbits. This can be done by exploring the geodesics and solving the geodesic equations.  In \cite{Kuniyal:2015uta, Soroushfar:2016yea} the geodesics of the magnetically charged Garfinkle-Horowitz-Strominger (GHS) stringy BH \cite{Garfinkle:1990qj} were studied and the absence of stable circular orbits outside the event horizon was found for massless test particles. In the background of a magnetically charged GHS BH the motion of massive particles with electric and magnetic charges was investigated, and bound and unbound orbits were found depending on critical values of the BH magnetic charge and the magnetic charge of the test particle.

All possible trajectories around Euler-Heisenberg (EH) BHs were studied in \cite{Amaro:2020xro}. By studying the geodesic equations and the corresponding effective potentials, it was found that the stable and unstable circular orbits of massive test particles are barely influenced by the presence of the non-linear EH electromagnetic field.  However, in the case of EH AdS BH the Lyapunov exponent was studied in \cite{Chen:2022tbb} and it was found that chaotic bounds were found and fixing  the particle charge and changing its angular momentum these bounds were violated.

A detailed study of null geodesics was carried out in \cite{Cardoso:2008bp}. By computing the Lyapunov exponent, which is the inverse of the instability timescale associated with a geodesic motion, it was shown that in a Myers-Perry black hole background spacetime \cite{Myers:1986un} with dimensions greater than four, the equatorial circular timelike geodesics are unstable, and the instability timescale of equatorial null geodesics in Myers-Perry spacetimes has a local minimum for spacetimes of dimension $d\geq6$. One of the reasons for this study was that in \cite{Pretorius:2007jn} some evidence was discussed that unstable circular orbits could give information on phenomena occurring at the threshold of black hole formation in high-energy scattering of black holes. This is obvious that this process  is of great   interest in fundamental physics as was discussed in \cite{Sperhake:2008ga,Shibata:2008rq}.

The aim of this work is to study the motion of massless particles in the background of one of the most well studied  scalar-tensor theory of gravity, the Horndeski theory. In this gravity theory, the scalar field is kinetically coupled to the Einstein tensor, and we study the null geodesic structure for asymptotically flat, AdS, and dS in the background of a Horndeski black hole. Our main purpose is to study the effect of the presence of a cosmological constant on the orbits. Also, we consider three classical tests of gravity in the solar system, such as the bending of the light, the gravitational redshift, and the Shapiro time delay, to constraint the coupling parameters of the scalar field to gravity.  Calculating also the Lyapunov exponent, we explore the stability of these geodesics in the presence of a cosmological constant.  Recently, the geodesic structure of rotating spacetimes in scalar–tensor theories has attracted considerable attention, particularly in the context of Horndeski gravity. In this framework, rotating black hole solutions that deviate from the Kerr geometry can exhibit distinct phenomenological signatures due to the presence of nontrivial scalar fields. In a recent study, the orbital precession and Lense–Thirring effect around rotating Horndeski black holes were studied, and it was shown  how the scalar–gravitational coupling modifies the nodal and periastron precession frequencies of timelike orbits, as well as the gyroscopic precession \cite{Zhen:2025nah}.

The paper is organized as follows. In Section \ref{bhHorn} we give a brief review of the four-dimensional Horndeski black hole discussed in \cite{Babichev:2017guv} which is considered the background black hole.  In Section \ref{NGS}, we study the motion of massless particles and establish their geodesic structure. Then, in Section \ref{LE} by calculating the Lyapunov exponent we study the stability of the geodesic structure and finally in Section \ref{conclution} we conclude.

\section{Four-Dimensional Horndeski Black Hole}
\label{bhHorn}

In this section after reviewing the Horndeski theory we will discuss a particular hairy black hole solution \cite{Rinaldi:2012vy} generated by a scalar field non-minimally coupled to the Einstein tensor. The action of the Horndeski theory \cite{Horndeski:1974wa} is given by
\begin{equation}
\label{acth}
S=\int d^4x\sqrt{-g}(L_2+L_3+L_4+L_5-2\Lambda)~,
\end{equation}
where 
\begin{gather*}
L_2=G_2(\phi,X)~,\quad L_3=-G_3(\phi,X)\Box \phi~, 
\end{gather*}
\begin{gather*}
\quad L_4=G_4(\phi,X)\mathcal{R}+G_{4,X}\left[(\Box\phi)^2-(\nabla_\mu\nabla_\nu\phi)(\nabla^\mu\nabla^\nu\phi)\right]\,, \\
\end{gather*}
and
\begin{eqnarray}
\notag L_5&=&G_5(\phi,X)G_{\mu\nu}\nabla^\mu\nabla^\nu\phi-\frac{1}{6}G_{5,X}[(\Box\phi)^3 \\
\notag &&  -3(\Box\phi)(\nabla_\mu\nabla_\nu\phi)(\nabla^\mu\nabla^\nu\phi)\\
&& +2(\nabla^\mu\nabla_\alpha\phi)(\nabla^\alpha\nabla_\beta\phi)(\nabla^\beta\nabla_\mu\phi)]~.
\end{eqnarray}
Here $g=\det(g_{\mu \nu})$ with $g_{\mu \nu}$ the metric tensor, $\mathcal{R}$ and $G_{\mu\nu}$ denote the Ricci scalar and the Einstein tensor respectively and $\Lambda$ is the cosmological constant. The functions $G_i$ with $i=2,3,4,5$ are arbitrary functions of the scalar field $\phi$ and the kinetic term $X = - \frac{1}{2} \partial_{\mu} \phi \partial^{\mu} \phi$, while $\Box\phi=\nabla_\mu\nabla^\mu\phi$ is the scalar field in the d'Alembertian operator with the covariant derivative $\nabla_{\mu}$, further $G_{j,X} = \partial G_j/\partial X$ with $j=4,5$.
\\

The action (\ref{acth}) is the most general action of the Horndeski theory. By fixing the various parameters of this action various black hole solutions were generated. 
The purpose of our work is to study the motion of particles in  a scalar-tensor theory of gravity in which the scalar field is kinetically coupled to the Einstein tensor. Such a theory can be generated in the Horndeski theory (\ref{acth}) by a specific choice of the parameters. 
In \cite{Babichev:2017guv} such a theory was generated by the following choice of the parameters \begin{eqnarray}
    G_{2} = \eta X, \quad G_{4}=\zeta + \beta \sqrt{-X}, \quad \text{and} \quad G_{3}=0=G_{5},
\end{eqnarray}
where $\eta$ with $\beta$ are dimensionless positive parameters and $\zeta=1/2$. With this choice of parameters the general action takes the form
\begin{eqnarray}
 \notag   S &=& \int d^4 x \sqrt{-g} \Bigg(\left[\frac{1}{2}+\beta\sqrt{(\partial \phi)^2/2} \right]\mathcal{R}-\frac{\eta}{2}(\partial\phi)^2\\
&&    -\frac{\beta}{\sqrt{2(\partial\phi)^2}} \left[(\Box\phi)^2-(\nabla_\mu\nabla_\nu \phi)^2 \right]-2\Lambda \Bigg) \,.
\end{eqnarray}
As can be seen in this action there are two crucial parameters, the $\eta$ parameter which controls the kinetic strength of the scalar field and the $\beta$ parameter which indicates the strength of interaction of the scalar field to curvature.   

Then a spherically symmetric black hole solution was found in \cite{Babichev:2017guv} which the scalar field generates primary hair appearing in the metric controlled by the parameter $\gamma$.

For a metric of the form

\begin{equation}
\label{metric}
 ds^{2}=-f(r)\, dt^{2}+\frac{ dr^{2}}{f(r)}+r^{2}( d\theta^{2}+\sin^{2}\theta\, d\phi^{2})\,,
 \end{equation} 
the   lapse function $f(r)$  was found to be 
\begin{equation} 
    f(r)=1-\frac{2M}{r}-\frac{\gamma ^{2}}{r^{2}}-\frac{\Lambda}{3}r^{2}\quad \text{with} \quad\gamma = \frac{\beta}{\sqrt{\eta}}  \,\label{1.2}.
\end{equation}

From this lapse function and depending on the value of the cosmological constant $\Lambda$, we can study the location of the horizons by analyzing the three different configurations separately:
\begin{enumerate}
	\item Asymptotically flat Horndeski black hole $(\Lambda=0)$: The spacetime allows a unique horizon (the event horizon $\rho_{+}$), which is located at
	\begin{equation}
	\rho_{+}=M\left(  1+\sqrt {1+{\gamma^2\over M^2} }\right) ,
	\label{g2.1}
    \end{equation}

and
    \begin{equation}
	\rho_{2}=M\left(  1-\sqrt {1+{\gamma ^2\over M^2} }\right) ,
	\label{g2.1b}
	\end{equation}
is a negative solution.
	\item  Horndeski --anti-de Sitter black hole $(\Lambda=-\frac{3}{\ell^{2}}<0)$: The spacetime allows a unique horizon (the event horizon $r_+$), which must be the real positive solution to the quartic equation
	\begin{equation} r^{4}+\ell^{2}r^2-2M\ell^{2} r-\gamma^{2}\ell^{2}=0 
  \,.  \label{1.3}\end{equation}
	 Its solutions are 
	\begin{eqnarray}
	        r_{+}&=&\alpha_{\ell} +\sqrt{{M\ell^2 \over 2\,\alpha_{\ell}}-{\ell^2 \over 2}-\alpha_{\ell}^2}\,,\\
        	r_{2}&=&\alpha_{\ell} -\sqrt{{M\ell^2 \over 2\,\alpha_{\ell}}-{\ell^2 \over 2}-\alpha_{\ell}^2}\,,\\
	        r_{3}&=&-\alpha_{\ell} +\sqrt{-{M\ell^2 \over 2\,\alpha_{\ell}}-{\ell^2 \over 2}-\alpha_{\ell}^2}\,,\\
	        r_{4}&=&-\alpha_{\ell} -\sqrt{-{M\ell^2 \over 2\,\alpha_{\ell}}-{\ell^2 \over 2}-\alpha_{\ell}^2}\,,
	\end{eqnarray}
	where  
	\begin{eqnarray}
	\alpha_{\ell} &=&\sqrt{U_{\ell}\cosh \left[ \frac{1}{3}\cosh^{-1} \Xi_{\ell}\right]-\frac{\ell^2}{6}}\,,\\
	U_{\ell}&=&\sqrt{{\ell^4 \over 36}-{\gamma^2 \ell^2 \over 3}}\,, \\
	\Xi_{\ell}&=&\frac{54M^2\ell+36\gamma^2\ell+\ell^3}{(\ell^2-12\gamma^2)^{3/2}}\,.
	\end{eqnarray}
        Through the discriminant for quartic polynomials, we can find in general that this is negative for all non-zero real $\ell$, so these four roots are distinct; on the one hand $r_+$ and $r_2$ are positive and negative real roots, respectively, while $r_3$ and $r_4$ are the complex conjugates of each other. 
	\item  Horndeski --de Sitter black hole $(\Lambda>0)$: The spacetime allows two horizons (the event horizon $R_+$ and the cosmological horizon $R_{++}$), which are obtained from the quartic equation 
    \begin{equation} 
        r^{4}-\frac{3}{\Lambda}r^2+\frac{6M}{\Lambda}r+\frac{3\gamma ^2}{\Lambda}=0\,.
        \label{1.5}
     \end{equation} 
Its solutions are 
    \begin{eqnarray}
          R_{++}&=&\alpha_{\Lambda} +\sqrt{-{3M \over 2\Lambda\,\alpha_{\Lambda}}+{3 \over 2\Lambda}-\alpha_{\Lambda}^2}\,,\\
         R_{+}&=&\alpha_{\Lambda} -\sqrt{-{3M \over 2\Lambda\,\alpha_{\Lambda}}+{3 \over 2\Lambda}-\alpha_{\Lambda}^2}\,,\\
        R_{3}&=&-\alpha_{\Lambda} +\sqrt{{3M \over 2\Lambda\,\alpha_{\Lambda}}+{3 \over 2\Lambda}-\alpha_{\Lambda}^2}\,,\\
        R_{4}&=&-\alpha_{\Lambda} -\sqrt{{3M \over 2\Lambda\,\alpha_{\Lambda}}+{3 \over 2\Lambda}-\alpha_{\Lambda}^2}\,,
    \end{eqnarray}
    where  
    \begin{eqnarray}
        \alpha_{\Lambda} &=&\sqrt{U_{\Lambda}\cosh \left[ \frac{1}{3}\cosh^{-1} \Xi_{\Lambda}\right]+\frac{1}{2\Lambda}}\,,\\
        U_{\Lambda}&=&\sqrt{{1 \over 4\Lambda^2}+{\gamma^2  \over \Lambda}} \,,\\
        \Xi_{\Lambda}&=&\frac{18M^2\Lambda+12\gamma^2\Lambda-1}{(1+4\gamma^2 \Lambda)^{3/2}}\,.
    \end{eqnarray}
    The roots $R_{3}$  and $R_{4}$ are negative solutions.
\end{enumerate}

\section{Null Geodesic structure}
\label{NGS}

In order to study the motion of test particles in the background (\ref{metric})
we use the standard Lagrangian approach \cite{Chandrasekhar:579245,Cruz:2004ts,Villanueva:2018kem}.
The corresponding Lagrangian is

\begin{equation}
2\mathcal{L} =\-f(r)\, \dot{t}^{2}+%
\frac{\dot{r}^{2}}{ f(r) }%
+r^{2}\left( \dot{\theta}^{2}+\sin ^{2}\theta \,\dot{\phi}^{2}\right) =-m^2.
\label{g6}
\end{equation}%
Here, the dot refers to the derivative with respect to an affine parameter $\lambda $,
along the trajectory, and, by normalization, $m=0\,(1)$ for massless (massive)
particles.
Since $(t, \phi)$ are cyclic coordinates, their corresponding
conjugate momenta $(\Pi _{t}, \Pi _{\phi })$ are conserved
and given by

\begin{equation}\label{g7}
\Pi _{t}=-f(r)\, \dot{t}=-E\,
,\qquad
\Pi _{\phi }=r^{2}\sin ^{2}\theta \,\dot{\phi}=L\,.
\end{equation}%
Therefore, considering that the motion is performed in an invariant plane, which
we fix at $\theta =\pi/2$, we obtain the
following expressions

\begin{equation}\label{g8}
\dot{t}=\frac{E}{f(r) }\,,%
\qquad\dot{\phi}=\frac{L}{r^{2}}\,.
\end{equation}

These relations together with Eq.(\ref{g6}) allow us to obtain the following
differential equations

\begin{eqnarray}\label{g9}
    \left(\frac{dr}{d\tau}\right)^{2}&=&E^2 - V_{\text{eff}}(r) ,\\ \label{g10}
    \left(\frac{dr}{dt}\right)^{2}&=&\frac{f(r)^2}{E^2}
    \,\left(E^2 - V_{\text{eff}}(r) \right), \\ \label{g11}
    \left(\frac{dr}{d\phi}\right)^{2}&=&\frac{r^{4}}{L^2}\left( E^2-V_{\text{eff}}(r) \right),
\end{eqnarray}%
where the effective potential $V_{\text{eff}}\left( r\right) $, reads

\begin{equation}
V_{\text{eff}}\left( r\right) =\left( 1-\frac{2M}{r}-\frac{\gamma ^{2}}{r^{2}}-\frac{\Lambda}{3}r^{2}\right)
\left( m^2+\frac{L^{2}}{r^{2}}\right) .  \label{g12}
\end{equation}%
In the next sections, based on this effective potential, we analyze the
null geodesic structure ($m=0$) of the space-time characterized by the metric (\ref{metric}).

\subsection{Motion with $L=0$}
\begin{figure}[H]
	\begin{center}
		\includegraphics[width=70mm]{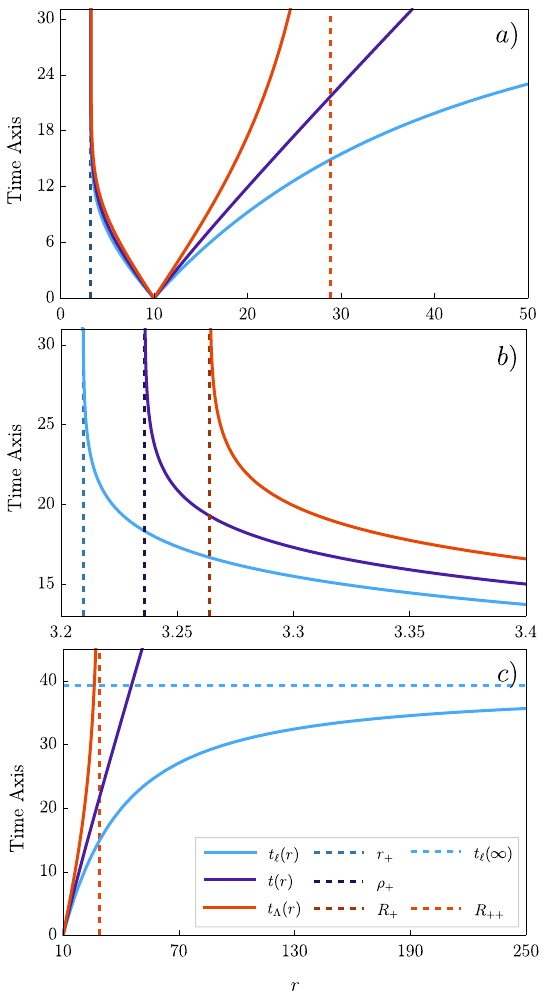}
	\end{center}
	\caption{Plot of the radial motion of massless particles.  We have considered $M=1$, $\Lambda=0.003$, $\ell=30$, $\gamma=2$, $\rho_0=10$, so $\rho_+=3.236$, $r_+=3.209$, $R_+=3.264$, $R_{++}=28.867$, and $t_{\ell}(\infty)=39.246$.}
	\label{f22}
\end{figure}

The motion of photons with vanishing angular momentum ($L=0$) is described by a null
 effective potential ($V_{\text{eff}}=0$) 
and therefore the photons in this case can escape to spatial infinity or plunged to the horizon.
Thus, the radial Eq. (\ref{g9}) reduces
to

\begin{equation}\label{n3}
\frac{dr}{d\lambda}=\pm E\,,
\end{equation}%
so, an elemental integration yields
\begin{equation}\label{n4}
\lambda\left( r,E\right) =\pm \frac{r-\rho_{0}}{E}\,.
\end{equation}%
Here, $\rho_{0}$ denotes the initial radial position of the massless particle. Note that the motion depends on the energy of the massless particle, and does not depend on the metric; thereby the above equation is valid for the three cases considered, that is, asymptotically flat, dS and AdS Horndeski spacetimes. 
\begin{enumerate}
	\item Asymptotically flat  Horndeski black hole  $(\Lambda=0)$:
by integrating Eq. (\ref{g10}) the coordinate time  yields 

\begin{eqnarray}\label{n5}
\notag t\left( r\right) &=& \pm
\Bigg[ r-\rho_0+ \frac{\rho^2_+}{\rho_{+}-\rho_{2}}
\ln \left|  \frac{r-\rho_{+} }{\rho_{0}-\rho_{+} }\right|\\
&& -
\frac{\rho^2_2}{\rho_{+}-\rho_{2}}
\ln \left|  \frac{r-\rho_{2}}{\rho_{0}-\rho_{2}}\right|
\Bigg]\,.
\end{eqnarray}%

In Fig. \ref{f1}, we show the affine parameter $\lambda$ and the coordinate time, for a massless particle. We observe that with respect to the
affine parameter the photons arrive at the horizon in a finite
affine parameter, and when the photons move in the opposite
direction, they require an infinity affine parameter to arrive
to infinity. Also, an observer located at $\rho_0$ will measure an infinite coordinate time for the photon to reach the event horizon. However, when the test particles move in the opposite direction, they require a infinite
coordinate time for arrive to infinity.  Note that when the energy of the photon increases the affine parameter decreases. \\

\begin{figure}[H]
	\begin{center}
		\includegraphics[width=8 cm]{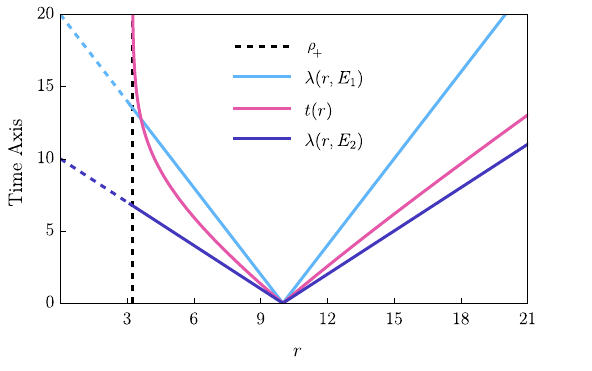}
	\end{center}
	\caption{Plot of the radial motion of massless particles.
        We have considered $M=1$, $\Lambda =0$, $\gamma=2$, $\rho_+=3.236$, $\rho_0=10$, $E_1=0.5$, and $E_2=1$.}
	\label{f1}
\end{figure}


	\item  Horndeski --de Sitter case $(\Lambda>0)$: the
	 expression for the coordinate time is 
	 \begin{equation}\label{nb5}
	 t_{\Lambda}\left( r\right) =\pm{3\over \Lambda}\sum_{i=1}^{4} F_i(r)
	 \end{equation}%
	 where
     \begin{widetext}
	 \begin{eqnarray}\label{Fi}
	F_1(r)&=&{-R_{++}^2\over (R_{++}-R_{3})(R_{++}-R_{4})(R_{++}-R_{+})}\ln \left|  \frac{R_{++}-r}{R_{++}-\rho_{0}}\right|,\\ 
	F_2(r)&=&{R_{+}^2\over (R_{+}-R_{3})(R_{+}-R_{4})(R_{++}-R_{+})}\ln \left|  \frac{r-R_{+}}{\rho_{0}-R_{+}}\right| ,\\ 
	F_3(r)&=&{-R_{3}^2\over (R_{3}-R_{4})(R_{++}-R_{3})(R_{+}-R_{3})}\ln \left|  \frac{r-R_{3}}{\rho_{0}-R_{3}}\right|,\\ 
	F_4(r)&=&{R_{4}^2\over (R_{3}-R_{4})(R_{++}-R_{4})(R_{+}-R_{4})}\ln \left|  \frac{r-R_{4}}{\rho_{0}-R_{4}}\right|\,.
	 \end{eqnarray}%
\end{widetext}

\item  Horndeski --Anti-de Sitter $(\Lambda<0)$: in this case the coordinate time is given by 
\begin{equation}\label{nb6}
t_{\ell}\left( r\right) =\pm{\ell^2}\sum_{i=1}^{4} \left( G_i(r)- G_i(\rho_{0})\right) 
\end{equation}%
where
\begin{widetext}
\begin{eqnarray}\label{Gi}
G_1(r)&=&\frac{{r_+}^2 \ln (r-{r_+})}{({r_+}-{r_2}) ({r_3} {r_4}+{r_+} ({r_+}-{r_3}-{r_4}))},\\ 
G_2(r)&=&-\frac{{r_2}^2 \ln (r-{r_2})}{({r_+}-{r_2}) ({r_3} {r_4}+{r_2} ({r_2}-{r_3}-{r_4}))},\\ 
G_3(r)&=&-\frac{ ({r_3} {r_4} ({r_2}+{r_+})-{r_2} {r_+} ({r_3}+{r_4}))\ln \left(r^2-r ({r_3}+{r_4})+{r_3} {r_4}\right)}{2 ({r_3}{r_4}+{r_2} ({r_2}-{r_3}-{r_4})) ({r_3} {r_4}+{r_+} ({r_+}-{r_3}-{r_4}))},\\ 
\notag G_4(r)&=&\frac{\left(2 {r_3}^2 {r_4}^2-{r_3} {r_4} (({r_2}+{r_+}) ({r_3}+{r_4})+2 {r_2} {r_+})+{r_2} {r_+} ({r_3}+{r_4})^2\right)}{\sqrt{4 {r_3} {r_4}-({r_3}+{r_4})^2} ({r_3} {r_4}+{r_2} ({r_2}-{r_3}-{r_4})) ({r_3} {r_4}+{r_+} ({r_+}-{r_3}-{r_4}))}\times\\
&\times &\tan ^{-1}\left(\frac{2 r-{r_3}-{r_4}}{\sqrt{4 {r_3} {r_4}-({r_3}+{r_4})^2}}\right) \,.
\end{eqnarray}%
\end{widetext}

In Fig. \ref{f22}  we show the coordinate time, for a massless particle for the three cases considered for a fixed value of $M$, and $\gamma$. Note that for the asymptotically flat case, as we mentioned, when the test particles move to the infinity, they require a infinite
coordinate time to arrive to infinity. However, when the spacetime is asymptotically AdS  the test particles require a finite
coordinate time to arrive to infinity given by $t_{\ell}\left( \infty\right)= \lim_{r\rightarrow \infty} t_{\ell}(r)$ which yields 
\begin{widetext}
\begin{eqnarray}\label{tinf}
\notag t_{\ell}\left( \infty\right)& =&{\ell^2}\left( \frac{\pi  \left(-{r_3} {r_4} (({r_2}+{r_H}) ({r_3}+{r_4})+2 {r_2} {r_H})+{r_2} {r_H} ({r_3}+{r_4}) ^2+2 {r_3}^2 {r_4}^2\right) }{2 \left(\sqrt{4{r_3} {r_4}-({r_3}+{r_4}) ^2} ({r_2} ({r_2}-{r_3}-{r_4})+{r_3}{r_4}) ({r_H} (-{r_3}-{r_4}+{r_H})+{r_3} {r_4})\right)}\right.\\ 
&-&\left. ({G_1}({\rho _0})+{G_2}({\rho_0})+{G_3}({\rho _0})+{G_4}({\rho_0}))\right)\,. 
\end{eqnarray}%
\end{widetext}
On the other hand, when the spacetime is asymptotically dS  the test particles require an infinity coordinate time in order to reach the cosmological horizon $R_{++}$. So, the effect of the cosmological constant is to introduce a distance limit for $\Lambda>0$, and a time limit for $\Lambda<0$.

\end{enumerate}

\clearpage

The asymptotic behavior of the coordinate time near horizons for different values of $\Lambda$ is described below

\begin{itemize}

\item For $ \Lambda = 0 $, we choose the negative sign in the solution (Eq.~34) to describe infall toward the event horizon. Then, the coordinate time near the event horizon is given by
\begin{equation}\label{t0}
t(r) \approx \frac{\rho_+^2}{\rho_+ - \rho_2}
\ln \left| \frac{\rho_0 - \rho_+}{r - \rho_+} \right|\,,
\end{equation}
from which it is clear that as $ r \rightarrow \rho_+ $, we have $ t \rightarrow \infty $.

\item  For $ \Lambda > 0 $, both an event horizon and a cosmological horizon exist. The infall toward the event horizon, described by Eq.~(35), can be written near the event horizon as
$
t(r) \approx -\frac{3}{\Lambda} F_2(r),
$
which yields
\begin{equation}\label{t1a}
t(r) \approx
\frac{3}{\Lambda} \frac{R_+^2}{(R_+ - R_3)(R_+ - R_4)(R_{++} - R_+)}
\ln \left| \frac{\rho_0 - R_+}{r - R_+} \right|\,,
\end{equation}
and again, as $ r \rightarrow R_+ $, it follows that $ t \rightarrow \infty $. Conversely, for the outgoing motion toward the cosmological horizon, also described by Eq.~(35), we have near the cosmological horizon
$
t(r) \approx \frac{3}{\Lambda} F_1(r),
$
leading to
\begin{equation}\label{t1b}
t(r) \approx
\frac{3}{\Lambda} \frac{R_{++}^2}{(R_{++} - R_3)(R_{++} - R_4)(R_{++} - R_+)}
\ln \left| \frac{R_{++} - \rho_0}{R_{++} - r} \right|\,,
\end{equation}
so that as $ r \rightarrow R_{++} $, the coordinate time diverges: $ t \rightarrow \infty $.

\item  For $ \Lambda < 0 $, we again choose the negative sign in the solution (Eq.~40) to describe infall toward the event horizon. In this case, the coordinate time near the event horizon is given by
$
t(r) \approx -\ell^2 \left( G_1(r) - G_1(\rho_0) \right),
$
which leads to
\begin{equation}\label{t2}
t(r) \approx \ell^2
\frac{r_+^2}{(r_+ - r_2)(r_3 r_4 + r_+ (r_+ - r_3 - r_4))}
\ln \left| \frac{\rho_0 - r_+}{r - r_+} \right|\,,
\end{equation}
indicating that as $ r \rightarrow r_+ $, the coordinate time diverges: $ t \rightarrow \infty $. Furthermore, since $ r_4 = r_3^* $, it follows that $ r_4 + r_3 \in \mathbb{R} $ and $ r_4 r_3 \in \mathbb{R} $, ensuring that the coordinate time remains real.

\end{itemize}


\subsection{Angular motion}

\begin{figure}[!h]
	\begin{center}
		\includegraphics[width=96mm]{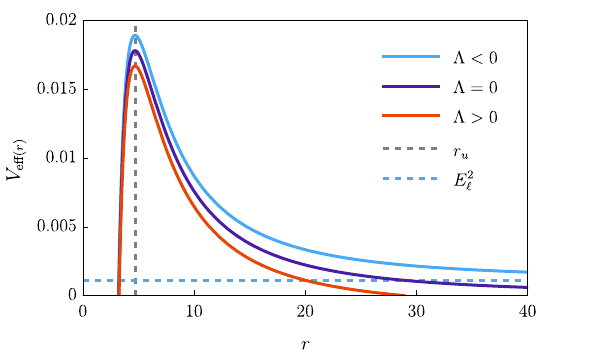}
	\end{center}
	\caption{Plot of the effective potential of photons. Here we have used the values $M=1$, $\Lambda=0.003$, $\ell=30$, $\gamma=2$, and $L=1$. The plot shows that the value of the instability distance is $r_u=4.702$, where the effective potential is maximum and it is independent of the cosmological constant. Also, $E^2_{\ell}= 0.001$.}
	\label{f2}
\end{figure}

Now, we study the motion on $L\neq 0$ with the effective potential $V_{\text{eff}}$ from (\ref{g12}) in the case $m=0$, namely
\begin{equation}\label{n1}
V_{\text{eff}}\left( r\right) =\frac{L^{2}}{r^{2}}-\frac{2ML^{2}}{r^{3}}-\frac{\gamma ^{2}L^{2}}{r^{4}}-
\frac{\Lambda L^{2}}{3}\,.
\end{equation}
A typical graph of this effective potential is shown in Fig. \ref{f2}. Now, in order to calculate the value of $r$, where the effective potential is maximum ($r_u$), we consider $V'_{eff}(r)$ given by 
\begin{equation}\label{Vp}
V'_{\text{eff}}\left( r\right) =-2\frac{L^{2}}{r^{3}}+\frac{6ML^{2}}{r^{4}}+\frac{4\gamma ^{2}L^{2}}{r^{5}}\,,
\end{equation}
and the condition $V'_{eff}(r) = 0 \quad\Rightarrow \quad  r^2-3M-2\gamma^2=0\,,
$
which corresponds to a quadratic equation, whose positive solution is
\begin{equation}
r_{u}=\frac{3M}{2}\left( 1+\sqrt{1+\frac{8\gamma^{2}}{9M^{2}}}\right)\,,
\label{rul}
\end{equation}
which represents an unstable circular orbit, that is independent of the cosmological constant. Note that evaluating the effective potential Eq. (\ref{n1}) at $r_{u}$, given by Eq. (\ref{rul}), we obtain
$E_{u}^{\ell}>E_{u}^0>E_{u}^{\Lambda}$, thereby $b_{u}^{\ell}<b_{u}^0<b_{u}^{\Lambda}$. Thus, the photon requires a greater energy for an AdS spacetime, which presents a smaller impact parameter,  in order to have this kind of orbit, and a lower energy for a dS spacetime, which presents a greater impact parameter. 

It is worth to mention that the presence of a cosmological constant modifies the energy threshold for the existence of unstable orbits, yet does not affect the radial distance $ r_u $ of such orbits. This asymmetry arises from the structure of the effective potential $ V_{\text{eff}}(r) $, where the cosmological constant contributes to the overall energy scale without shifting the extrema of the potential in certain regimes. A possible theoretical explanation is that the cosmological constant acts as a background energy density, modifying the total energy budget required for a particle to reach or maintain a particular orbit, but not necessarily influencing the curvature of the potential in a way that changes the location of unstable points.

\begin{figure}[H]
	\begin{center}
		\includegraphics[width=90mm]{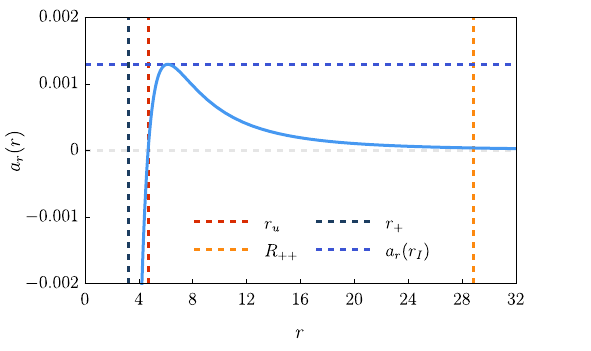}
	\end{center}
	\caption{Plot of the radial acceleration for massless particles. Here, $M=1$, $\gamma=2$, and $L=1$.
	The graph shows the radial acceleration, which is independent of the cosmological constant. The radial acceleration is maximum, at the inflection point, $r_I=6.163$, of the effective potential. Also, $a_r(r_I)=0.0013$, $r_u=4.702$, $r_+=3.209$, and $R_{++}=28.867$.}
	\label{f2b}
\end{figure}

On the other hand, in Fig. \ref{f2b} we plot the radial acceleration for the massless particles given by $ a_{r} \equiv \ddot{r} =-V_{\text{eff}}^{\prime}(r)/2$. Now, in order to calculate the value of $r$, where the radial acceleration is maximum ($r_I$), we consider $V''_{eff}(r)$ given by 
\begin{equation}\label{Vpp}
V''_{\text{eff}}\left( r\right) =6\frac{L^{2}}{r^{4}}-\frac{24ML^{2}}{r^{5}}-\frac{20\gamma ^{2}L^{2}}{r^{6}}\,,
\end{equation}
and the condition $V''_{eff}(r) = 0 \quad\Rightarrow \quad  r^2-4M-{10\over3}\gamma^2=0$, which corresponds to a quadratic equation whose positive solution is

\begin{equation}
r_{I}=2M\left( 1+\sqrt{1+\frac{5\gamma^{2}}{6M^{2}}}\right)\,.
\label{n8l}
\end{equation}
Also, note that for $r_+<r<r_u$, the  radial acceleration $a_{r}<0$, for $r=r_u$, the  radial acceleration $a_r=0$, for $r_u<0<\infty$, the  radial acceleration $a_r>0$.

Now, in order to perform a qualitative analysis of the effective potential, we will consider the impact parameter. So, considering Eq. (\ref{g11}), and (\ref{n1}), we obtain 
\begin{equation}\label{em1}
\left(\frac{dr}{d\phi}\right)^{2}=\left(\frac{1}{b^2}+ \frac{\Lambda }{3}\right)r^{4} -r^{2}+2Mr+\gamma ^{2}\,,
\end{equation}%
where $b \equiv L/E$ is the impact parameter.
Therefore, based on the impact parameter values and Fig. (\ref{f2}), we present a brief qualitative description of the allowed angular motions for photons in Horndeski black holes.

\begin{itemize}
	\item[(i)] {\it Capture zone:} 	If $0 < b < b_u$, the photons fall inexorably to the horizon $r_+$, or escape to infinity, depending on the initial conditions, and their cross section, $\sigma$, in this geometry is \cite{wald}
	\begin{equation}
	\sigma=\pi\,b_u^2={\pi\,r_u^2 \over f(r_u)}\,.
	\label{sigma}
	\end{equation}
	\item [(ii)] {\it Critical trajectories:} 
	If $b=b_u$, photons can stay in one of the unstable inner circular orbits of radius $r_u$. Therefore, photons that arrive from the initial distance $r_i$ ($r_+ < r_i < r_u$, or $r_u < r_i < \infty$) can asymptotically fall to a circle of radius $r_u$. The affine period in such orbit is	
		\begin{equation}
	T_{\lambda}={2\pi\,r_u^2 \over L}\,,
	\label{T1}
	\end{equation}
		and the coordinate period is
		\begin{equation}
		T_{t}=2\pi\,b_u\,.
		\label{T2}
		\end{equation}
	\item [(iii)] {\it Deflection zone:} If $b_u < b < b_{\ell}$, this zone presents orbits of the first and second kind. The orbits of the first kind are allowed in the interval $r_d \leq r < \infty$, where the photons can come from a finite distance or from an infinity distance until they reach the distance $r = r_d$ (which is a solution of the equation $V_{\text{eff}}(r_d) = E^2$), and then the photons are deflected. Note that photons with $b \geq b_{\ell}= {\ell}$ are not allowed in this zone. The orbits of the second kind are allowed in the interval $r_+ < r \leq r_ F$, where the photons come from a distance greater than the event horizon, then they reach the distance $r_F$ (which is a solution of the equation $V_{\text{eff}}(r_F) = E^2$) and then they plunge into the horizon.	
		\item [(iv)] {\it Second kind and lima\c con with loop geodesic} 
		If $b_u \leq b < \infty$, the return point is in the range $r_+ < r < r_u$, and then the photons plunge into the horizon. However, when $b = b_{\ell}$ a special geodesic can be obtained, known as the lima\c con with loop.	
	\end{itemize}

\subsubsection{Bending of light}
Now, in order to obtain the bending of light, we consider
Eq. (\ref{em1}), which can be written as
\begin{equation}	
\label{em}
\left(\frac{dr}{d\phi}\right)^{2}
={ r^{4}-\mathcal{B}^2 r^{2}+2M\mathcal{B}^2r+\mathcal{B}^2\gamma^2\over \mathcal{B}^2}\equiv {\mathcal{P}(r)\over \mathcal{B}^2}\,,
\end{equation}
where
\begin{eqnarray}
    \mathcal{B} = \left(\frac{1}{b^2}+ \frac{\Lambda }{3}\right)^{-1/2}
\end{eqnarray}
is the {\it anomalous impact parameter},  defined as the combination of the usual impact parameter $ 1/b^2 $ and the cosmological constant $ \Lambda $.

\begin{itemize}
\item  For $ \Lambda > 0 $, we always have $ \frac{1}{\mathcal{B}^2} > 0 $, and the null geodesics are qualitatively similar to those in the Schwarzschild case.

\item For $ \Lambda < 0 $, the expression becomes
\begin{equation}
\frac{1}{\mathcal{B}^2} = \frac{1}{b^2} - \frac{1}{\ell^2},
\end{equation}
where $ \ell $ is the AdS radius. This formulation is particularly relevant in AdS spacetimes, as it allows for the characterization of a new region of spacetime that is not permitted in the $ \Lambda \geq 0 $ cases. When $ \frac{1}{\mathcal{B}^2} \leq 0 $, new second-class trajectories emerge, associated with energies in the range $ 0 < E \leq E_\ell $. In the special case where $ \frac{1}{\mathcal{B}^2} = 0 $, the second-class trajectory corresponds to a limaçon with loop. This new region, along with the appearance of exotic geodesics, is a distinctive feature of spherically symmetric AdS spacetimes \cite{Villanueva:2013zta,Gonzalez:2020zfd}.
\end{itemize}

Now, in order to obtain the return points, we solve the equation $\mathcal{P}(r ) = 0$,  so, the turning point is located at%
\begin{eqnarray}
    r_{d}&=&\alpha +\sqrt{{\mathcal{B}^2 \over 2}-\alpha^2-{M\mathcal{B}^2 \over 2\,\alpha}}\,,\\
    r_{f}&=&\alpha -\sqrt{{\mathcal{B}^2 \over 2}-\alpha^2-{M\mathcal{B}^2 \over 2\,\alpha}}\,,\\
    r_{3}&=&-\alpha +\sqrt{{\mathcal{B}^2 \over 2}-\alpha^2+{M\mathcal{B}^2 \over 2\,\alpha}}\,,\\
    r_{4}&=&-\alpha -\sqrt{{\mathcal{B}^2 \over 2}-\alpha^2+{M\mathcal{B}^2 \over 2\,\alpha}}\,,
\end{eqnarray}
where
\begin{eqnarray}
    \alpha &=&\sqrt{U_0\cosh \left[ \frac{1}{3}\cosh^{-1} \Xi_0\right]+\frac{\mathcal{B}^2}{6}}\,,\\
    U_0&=&\sqrt{{\mathcal{B}^4 \over 36}+{\gamma^2 \mathcal{B}^2 \over 3}}\,, \\
    \Xi_0&=&\frac{54M^2\mathcal{B}+36\gamma^2\mathcal{B}-\mathcal{B}^3}{(\mathcal{B}^2+12\gamma^2)^{3/2}}\,,
\end{eqnarray}
$ r_d$ is the deflection distance, $ r_f$ is the return point and $ r_3$ with $ r_4$ are negative solutions. Now, using Eq. (\ref{em}), we obtain the quadrature
\begin{equation}\label{n9}
\phi \left( r\right) =\mathcal{B}\int_{r_{d}}^{r}\frac{\,dr'}{\sqrt{\left( r'-r_{d}\right)
		\left( r'-r_{f}\right) \left( r'-r_{3}\right) \left( r'-r_{4}\right) }}\,.
\end{equation}%

In order to integrate out (\ref{n9}) it is instructive
to make the change of variable $r=r_{d}\left( 1+{1\over 4U-\alpha_d/3}\right) $, and
after a brief manipulation, we obtain the polar
trajectory of massless particles
\begin{equation}\label{n11}
r\left( \phi \right) =r_{d}+{r_{d}\over 4\wp \left(\kappa_d\,\phi
	;g_{2},g_{3}\right)-\alpha_d/3}\,,
\end{equation}%
where 
\begin{eqnarray}
    \kappa_{d}&=&{r_d\over \mathcal{B}\sqrt{u_4u_3u_f}}, \\
    \alpha_{d}&=&u_4+u_3+u_f,
\end{eqnarray}
while $\wp \equiv \wp(y; g_2, g_3)$ is the $\wp$-Weierstra{\ss} function and
$g_{2}$ with $g_{3}$ are the so-called
Weierstra{\ss} invariants, given by

\begin{eqnarray}
\notag  g_{2}&=&\frac{1}{12}\left(  u_{4}^2+u_{3}^2+u_{f}^2-u_{4}\,u_{3}-u_{4}\,u_{f}-u_{f}\,u_{3}\right)\,,\\
\notag g_{3}&=&\frac{1}{432}\left(2 u_{4}-u_{3}-u_{f}  \right) \left(2 u_{3}-u_{4}-u_{f}  \right)\cdot \\
&&\left( u_{4}+u_{3}-2u_{f}  \right) \,.
\end{eqnarray}

\begin{figure}[!h]
	\begin{center}
		\includegraphics[width=65mm]{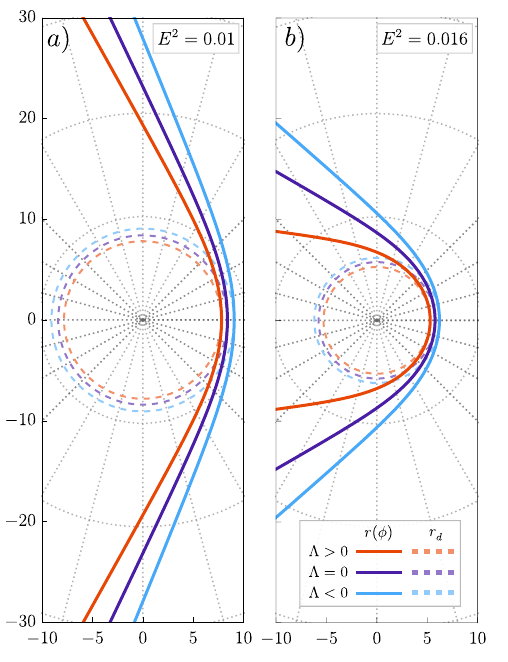}
	\end{center}
	\caption{Polar plot for deflection of light with $M=1$, $\Lambda=0.003$, $\ell=30$, $\gamma=2$, and $L = 1$. All trajectories have the same energy $E^2=0.01$ and $E^2=0.016$, respectively.}
	\label{f3}
\end{figure}

The other constants are: $u_{4}={r_d\over r_d-r_4}$, $ u_{3}={r_d\over r_d-r_3}$ and $u_{f}={r_d\over r_d-r_f}$. In Fig. \ref{f3} we show the behavior of the bending of light given by Eq. (\ref{n11}), for a fixed value of the black hole mass $M$, the angular momentum $L$, and the parameter $\gamma$, we observe that the deflection angle is greater when the black hole has a positive cosmological constant. On the other hand, note that the deflection distance, $r_d$, is greater when the black hole has a negative cosmological constant. 
\bigskip


It is well known that photons can escape to infinity during a scattering process. So, considering $r(\phi)|_{\phi=0} = r_d$, the shortest distance to the black hole at which the deflection happens, and assuming that the incident photons come from infinity and escape to infinity, we have $r(\phi)|_{\phi_{\infty}} = \infty$. Now, by using Eq. (\ref{n11}) we obtain that the deflection angle, $\hat{\alpha}=2|\phi_{\infty}|-\pi$, is given by
\begin{equation}\label{angdef}
\hat{\alpha}={2\over \kappa_d} \left| \wp^{-1}\left({\alpha_d\over 12} 	;g_{2},g_{3}\right) \right| -\pi\,.
\end{equation}

\begin{figure}[!h]
	\begin{center}
		\includegraphics[width=90 mm]{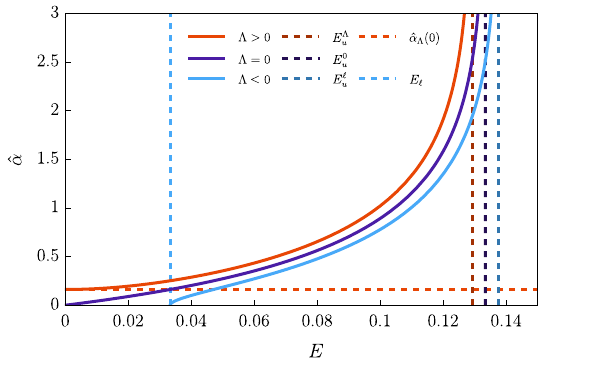}
	\end{center}
	\caption{The behavior of the deflection angle $\hat{\alpha}$ in terms of $E$, with  $M=1$, $\Lambda=0.003$, $\ell=30$, $\gamma=2$ and $L = 1$. $E_{u}^{\ell}=0.138$, $E_{u}^0=0.133$, $E_{u}^{\Lambda}=0.129$, $E_{\ell}=0.033$, and $\hat{\alpha}_{\Lambda}(0)=0.162$.}
	\label{defleccion2}
\end{figure}

The evolution of the deflection angle has been plotted in Fig. \ref{defleccion2}, which shows an asymptotic behavior as $E\rightarrow E_u$, for a fixed value of the black hole mass $M$, the angular momentum $L$, and the parameter $\gamma$, we can observe that the deflection angle takes an infinite value when $E = E_u$, such that $E_u$  is greater, when the black hole has a negative cosmological constant. When $E\rightarrow 0$, the deflection angle is null for asymptotically flat, but when the cosmological constant is positive (dS), the deflection angle is finite and not null $\hat{\alpha}_{\Lambda}(0)$. In the case of an AdS spacetime, when $E\rightarrow E_{\ell}$, the deflection angle is null. Also, note that for a fixed value of the deflection angle, the particle requires a greater energy when the spacetime is AdS.

\bigskip

The process of light deflection, or the so-called gravitational lensing, can theoretically be approached by means of the geodesic equations for the light rays (null geodesics). 
Accordingly, the first-order, angular equation of motion for the light rays (i.e. photons as test particles) passing near the black hole is given by \eqref{em1}.
Performing the change of variable $r=1/u$, the above equation yields
\begin{equation}\label{ue}
\left(\frac{ du}{d\phi}\right)^2=\frac{1}{b^2}+ \frac{\Lambda }{3} -u^{2}+2Mu^3+\gamma ^{2}u^4\,,
\end{equation}
that reduces to the standard Schwarzschild equation of light deflection in the limit of $\Lambda \rightarrow 0$ and $\gamma 	\rightarrow  0$.
Differentiating Eq. \eqref{ue} with respect to $\phi$, gives
\begin{equation}
u^{\prime\prime}+ u =3Mu^2+2\gamma ^{2}u^3,
\end{equation}
where the prime denote differentiations with respect to $\phi$. Following the procedure established in Ref. \cite{Straumann}, we obtain
\begin{equation}
u=\frac{1}{b}\sin\phi+{ 3M\over 2\,b^2}+\frac{\sqrt{2}\,\gamma^2}{2\,b^3}
+\left({ M\over 2\,b^2}+\frac{\sqrt{2}\,\gamma^2}{4\,b^3}\right)\cos(2\phi)\,.
\end{equation}
Note that, $u \rightarrow 0$ results in $\phi\rightarrow\phi_{\infty}$, with
\begin{equation}
-\phi_{\infty}={2M\over b}+\frac{3\sqrt{2}\,\gamma^2}{4\,b^2}\,.
\end{equation}
The deflection angle of the light rays passing the black hole is, therefore, obtained as
\begin{equation}\label{GB1}
\hat{\alpha} = 2\left |-\phi_{\infty}\right | = {4M\over b}+\frac{3\sqrt{2}\,\gamma^2}{2\,b^2}\,,
\end{equation}
which recovers the famous form of $\hat{\alpha}_{\mathrm{Sch}}=4M/b$ for the Schwarzschild black hole in the limit  $\gamma\rightarrow  0$. Also, note that the constant $\Lambda$ has no influence on the light deflection, as was pointed out in Ref. \cite{Kagramanova:2006ax}. Assuming the Sun as the central massive object, the deflection angle predicted by the Schwarzschild solution is given by
$\hat{\alpha}_{\mathrm{Sch}} = \frac{4M_{\odot}}{R_{\odot}}$
which, in arcsecond, reads
$
\hat{\delta}_{\mathrm{Sch}} = \frac{3600 \cdot 180}{\pi} \hat{\alpha}_{\mathrm{Sch}}=1.75092 \, \text{arcsec}\,. 
$
Observations near the Sun report measured values of
$\hat{\alpha}_P = 1.7520\, \text{arcsec}$ (prograde) and $\hat{\alpha}_R = 1.7519\, \text{arcsec}$ (retrograde)~\cite{Roy:2019ijp,Fathi:2025byw}. Therefore, by identifying the theoretical prediction with the observational value, we obtain the following constraints:
\begin{align}
\hat{\delta}_{\mathrm{SH}} = \hat{\alpha}_P &\quad \Rightarrow \quad \gamma = 34541\, \text{m}, \\
\hat{\delta}_{\mathrm{SH}} = \hat{\alpha}_R &\quad \Rightarrow \quad \gamma = 32899\, \text{m}.
\end{align}



\subsubsection{Second kind trajectories and  lima\c con with loop geodesic}

The spacetime allows second kind trajectories, when $b_u < b<\infty$, where the turn point is in the range $r_+ <r <r_u$, and then the photons plunge into the horizon. 
In order to integrate out (\ref{n9}) from the initial distance $r_f$ to $r$, it is useful
to make the change of variable $r=r_{f}\left( 1-{1\over 4U-\alpha_f/3}\right) $, and
after a brief manipulation, we obtain the polar
trajectory of massless particles
\begin{equation}\label{n111}
r\left( \phi \right) =r_{f}-{r_{f}\over 4\wp \left(\kappa_f\,\phi
	;g_{2f},g_{3f}\right)-\alpha_f/3}\,,
\end{equation}%
where 
\begin{eqnarray}
    \kappa_{f}&=&\frac{r_f}{ \mathcal{B}\sqrt{u_{4f}u_{3f}u_d}} \\
    \alpha_{f}&=&u_d-u_{3f}-u_{4f},
\end{eqnarray}
while $\wp \equiv \wp(y; g_{2f}, g_{3f})$ is the $\wp$-Weierstra{\ss} function and
$g_{2f}$ with $g_{3f}$ are the so-called
Weierstra{\ss} invariants, given by

\begin{eqnarray}
 \notag g_{2f}&=&\frac{1}{12}\left(  u_{4f}^2+u_{3f}^2+u_{d}^2-u_{4f}\,u_{3f}+u_{4f}\,u_{d}+u_{d}\,u_{3f}\right)\,,\\
\notag g_{3f}&=&\frac{1}{432}\left(u_{3f}-u_{d}-2 u_{4f}  \right) \left(2 u_{3f}-u_{4f}+u_{d}  \right)\cdot \\
&& \left( u_{4f}+u_{3f}+2u_{d}  \right)\,. 
\end{eqnarray}

\begin{figure}[!h]
	\begin{center}
		\includegraphics[width=80mm]{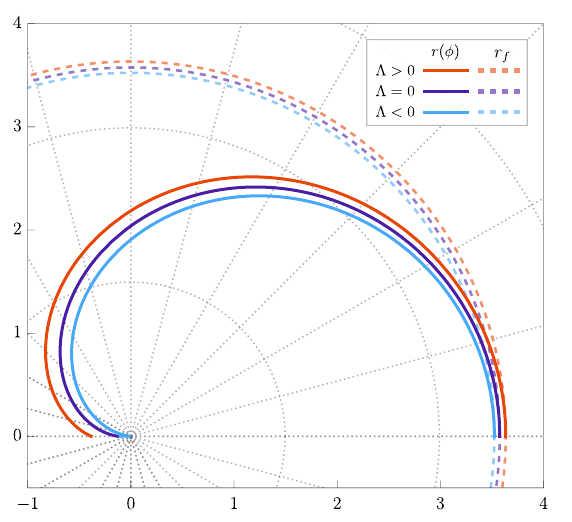}
	\end{center}
	\caption{Polar plot for second kind trajectories with $M=1$, $\Lambda=0.003$, $\ell=30$, $\gamma=2$, and $L = 1$. All trajectories have the same energy $E^2=0.01$.}
	\label{caida}
\end{figure}

The other constants are: $u_{4f}={r_f\over r_f-r_4}$, $ u_{3f}={r_f\over r_f-r_3}$ and $u_{d}={r_f\over r_d-r_f}$. In Fig. \ref{caida} we show the behavior of the second kind trajectories, given by Eq. (\ref{n111}), for a fixed value of the black hole mass $M$, the angular momentum $L$, the energy $E$ and the parameter $\gamma$, we observe that the initial distance $r_f$ is greater when the spacetime is dS.

\bigskip

\begin{figure}[!h]
	\begin{center}
		\includegraphics[width=70mm]{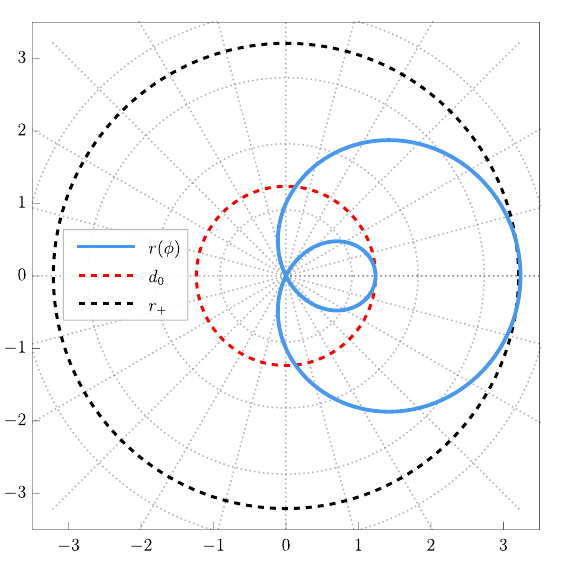}
	\end{center}
	\caption{The lima\c con with loop  geodesic  (light blue line), with $E_{\ell}= 0.033$, dashed black lines correspond to the event horizon and dashed red lines is $d_0$. Here,  $M = 1$,  $\ell=30$, $\gamma=2$, $L = 1$, $r_0=3.236$, and $r_+=3.209$.}
	\label{limazon3}
\end{figure}

The AdS spacetime allows second kind trajectories, when $b_{\ell} < b<\infty$, where the turn point is in the range $r_+ <r <r_0$, and then the photons plunge into the horizon. However, there is a special geodesic for AdS spacetime, which can be obtained when the anomalous impact parameter is $\mathcal{B }\rightarrow\infty \, (b = \ell)$. In this case, the radial coordinate is restricted to $r_+ < r < r_0$, and the equation of motion (\ref{em1}) can be written as

\begin{equation}\label{lima1}
\phi(r)=-\int_{r_0}^{r} {dr'\over  \sqrt{-r'^{2}+2Mr'+\gamma ^{2}}}  \,,        
\end{equation}%
and the return points are
\begin{eqnarray}\label{lima2}
r_{0}&=&M+\sqrt{M^2+\gamma^2}\,,\\
d_{0}&=&M-\sqrt{M^2+\gamma^2}\,.
\end{eqnarray}
Thus, it is straightforward to find the solution of Eq. (\ref{lima1}), which is given by
\begin{equation}\label{lima3}
r(\phi)=M+   \sqrt{M^{2}+\gamma ^{2}}\,\cos \phi  \,, 
\end{equation}%
which represents the lima\c con with loop  geodesic  of Pascal, see Fig. \ref{limazon3}. This trajectory is a new type of orbit in Hordeski AdS, and it does not depend on the value of the cosmological constant. 

\bigskip

It is worth mentioning that the analog geodesic in four-dimensional RN AdS corresponds to the lima\c con of Pascal \cite{Villanueva:2013zta}. Also, when the spacetime is the five-dimensional Schwarzschild–anti-de Sitter spacetime this geodesic is given by $r = 2 M \cos \phi$, which describes a circumference with radius $M$ that is analogous to the cardioid geodesics found in four-dimensional Schwarzschild–anti-de Sitter spacetime \cite{Cruz:2004ts}, and the Hippopede of Proclus geodesic when the spacetime is five-dimensional RNAdS \cite{Gonzalez:2020zfd}.

The limaçon with loop geodesic belongs to the null geodesic structure, within which light deflection stands as the most significant observational evidence. Other geodesics, such as critical trajectories and second class trajectories, which complete the structure, remain unobservable at present. The limaçon with loop, together with other exotic geodesics mentioned such as the cardioid, Pascal’s limaçon, and Proclus’s hippopede, arise in spherically symmetric anti-de Sitter (AdS) spacetimes. Consequently, their physical relevance is closely tied to the existence of a negative cosmological constant. The existence of a negative cosmological constant (i.e., an anti–de Sitter-type vacuum in the dark energy sector), complemented by a quintessence field have been studied, for instance in Refs. \cite{Sen:2021wld, Menci:2024rbq}, which suggest that a 
$\Lambda<0$, which can naturally arise in string theory frameworks, is not ruled out by current cosmological observations.

\subsubsection{Critical trajectories}

\begin{figure}[!h]
	\begin{center}
		\includegraphics[width=80mm]{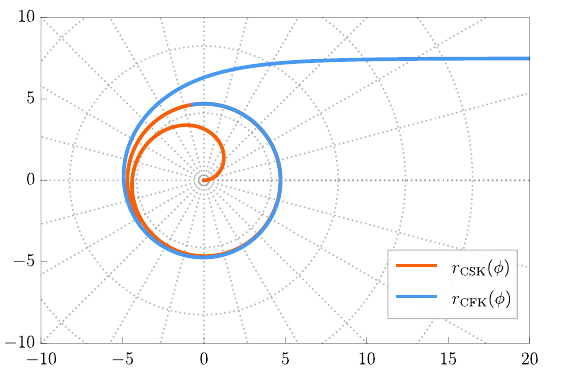}
	\end{center}
	\caption{The critical trajectories  plotted for $M = 1$,  $\gamma=2$ and $L = 1$, with  $r_u =4.7$. Blue  line for CFK and red  line for CSK trajectories.}
	\label{criticas}
\end{figure}

In the case of $b = b_u$, the particles can be confined on unstable circular orbits of radius $r_u$. This kind of motion is indeed ramified into two cases; critical trajectories of the first kind (CFK) in which the particles come from a distant position $r_i$ to $r_u$ ($r_i > r_u$) and those of the second kind (CSK) in which the particles start from an initial point $d_i$ in the vicinity of $r_u$ ($d_i < r_u$) and then tend to this radius by spiraling. We obtain the following equations of motion for the aforementioned trajectories:

\begin{equation}\label{criti1}
r_{\text{CFK}}(\phi)=r_u+{2(r_u-r_3)(r_u-r_4)\over (r_3-r_4)\cosh(\kappa_u\,\phi+\varphi_{\infty})+r_3+r_4-2r_u} \,,
\end{equation}%
where 
\begin{eqnarray}
    \kappa_u&=&{\sqrt{(r_u-r_3)(r_u-r_4)}\over\mathcal{ B}_u}, \\
    \varphi_{\infty}&=&\cosh^{-1}\left({2r_u-r_3-r_4\over r_3-r_4 } \right),
\end{eqnarray}
for the first kind, while for the second kind  
 \begin{equation}\label{criti2}
 r_{\text{CSK}}(\phi)=r_u-{2(r_u-r_3)(r_u-r_4)\over (r_3-r_4)\cosh(\kappa_u\,\phi+\varphi_{0})-r_3-r_4+2r_u} \,,
 \end{equation}%
 where  
 \begin{eqnarray}
    \varphi_{0}=\cosh^{-1}\left({ r_3(r_4-r_u) + r_4(r_3-r_u) \over r_u(r_3-r_4) } \right). 
 \end{eqnarray}
In Fig. \ref{criticas}, we show the behavior of the CFK Eq. (\ref{criti1}) and the CSK Eq. (\ref{criti2}) trajectories. As we mentioned, $r_u$ is independent of the cosmological constant, and thus these trajectories are the same for asymptotically flat, dS and AdS spacetime. 

\bigskip


\subsubsection{Capture zone}
\begin{figure}[H]
	\begin{center}
		\includegraphics[width=80mm]{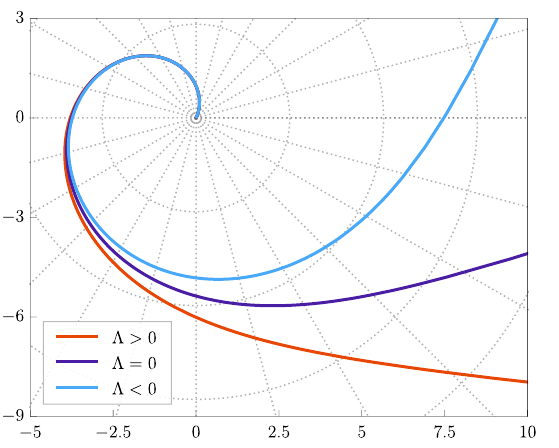}
	\end{center}
	\caption{The capture zone, trajectories  can plunge into the horizon or escape to infinity. Here, $M = 1$,  $\ell=30$, $\gamma=2$, $L = 1$, and $b=7.143$.}
	\label{captura}
\end{figure}

The photons with an impact parameter smaller than the critical one ($b<b_u$), which are in the capture zone, can plunge into the horizon or escape to infinity, with a cross section given by Eq. (\ref{sigma}). So, by manipulating Eq. (\ref{n11}), and considering $b<b_u$, we show the trajectories of the particles in the capture zone in Fig. \ref{captura}. 

\bigskip

\subsection{Gravitational redshift}

Since Horndeski black hole is a stationary spacetime there is a time-like Killing vector so that in coordinates adapted to the symmetry the ratio of the measured frequency of a light ray crossing different positions is given by \cite{Kagramanova:2006ax} 
\begin{equation}
{\nu \over \nu_0}=\sqrt{\frac{g_{00}(r)}{g_{00}(r_0)}}\,,
\end{equation}
for  $M/(r) \ll 1$ and $\gamma/r\ll1$, the above expression yields 
\begin{equation}
{\nu \over \nu_0}\approx  1+ M \left( \frac{1}{r_0}-\frac{1}{r}\right)+\frac{\gamma^2}{2}\left(\frac{1}{r_0^2}-\frac{1}{r^2}\right)+\frac{\Lambda}{6}\left(r_0^2-r^2\right)\,,
\end{equation}
where we have neglected products of $\gamma$ and $M$. Obviously, if we consider the limit ${M} 	\rightarrow M_{\oplus}$,  and $\gamma 	\rightarrow  0$, we recover the classical result for the Schwarzschild dS and AdS spacetime.
The clock can be compared with an accuracy of $10^{-15}$, the H-maser in the GP-A redshift experiment \cite{Vessot:1980zz} reached an accuracy of $10^{-14}$. Therefore, considering that all observations are well described within Einstein's theory with $\Lambda=0$, we conclude that the extra terms of Horndeski must be $<10^{-14}$. Thus,
\begin{equation}
\abs\gamma \leq 0.945\; \text{m}\,,
\end{equation}
where we assume a clock comparison between Earth and a satellite at 15,000 km height, as in Ref. \cite{Kagramanova:2006ax}, where the authors have constrained the cosmological constant.

\subsection{Shapiro time delay}

An interesting relativistic effect in the propagation of light rays is the apparent delay in the time of propagation for a light signal passing near the Sun, which is a relevant correction for astronomical observations and is called the Shapiro time delay. The time delay of Radar Echoes corresponds to the determination of the time delay of radar signals which are transmitted from the Earth through a region near the Sun to another planet or spacecraft and then reflected back to the Earth. The time interval between emission and return of a pulse as measured by a clock on Earth is \cite{Straumann}

\begin{equation}
t_{12}=2\, t(r_1,r_d)+2\, t(r_2,r_d)\,,
\end{equation}
where $r_d$ as closest approach to the Sun. Now, in order to calculate the time delay we use Eq. (\ref{g10}), 
and by considering that
$dr/dt$ vanishes, thereby
$\frac{E^2}{L^2}=\frac{f(r_d)}{r_d^2}$.
Thus, the coordinate time that the light requires to go from $r_d$ to $r$ is given by
\begin{equation}
t(r,r_d)=\int_{r_d}^r \frac{dr'}{f(r')\sqrt{1-\frac{r_d^2}{f(r_d)}\frac{f(r')}{r'^2}}}\,.
\end{equation}
So, at first order correction we obtain
\begin{equation}
t(r, r_d)=\sqrt{r^2-r_d^2}+t_M(r)+t_{\gamma}(r)+t_{\Lambda}(r)\,,
\end{equation}
where
\begin{eqnarray}
t_M(r)&=&M\left(\sqrt{r-r_d\over r+r_d}+2\,\ln\left| \frac{r+ \sqrt{r^2-r_d^2}}{r_d}\right|  \right) \,,\\
t_{\gamma}(r)&=&\frac{3\,\gamma^2}{2\,r_d}\sec^{-1}\left({r\over r_d}\right)\,,\\
t_{\Lambda}(r)&=&{\Lambda\over 18}\left(2\,r^2+r^2_d\right)\sqrt{r^2-r_d^2}\,.
\end{eqnarray}
Therefore, for the circuit from point 1 to point 2 and back the delay in the coordinate time is
\begin{equation}
\Delta t = 2\left[t(r_1, r_d)+t(r_2,r_d)-\sqrt{r_1^2-r_d^2}-\sqrt{r_2^2-r_d^2}\right]\,,
\end{equation}
namely,
\begin{eqnarray}
\notag
\Delta t &=&2\left[t_M(r_1)+t_M(r_2)+t_{\gamma}(r_1)+t_{\gamma}(r_2)\right]\\
&& +2\left[t_{\Lambda}(r_1)+t_{\Lambda}(r_2)\right]\,.
\end{eqnarray}

Now, for a round trip in the solar system, we have ($r_d <<r_1,r_2$)
\begin{eqnarray}
\label{deltat}
\notag
\Delta t &\approx& 4M\left(1+ \ln\left| \frac{4r_1r_2}{r_d^2}\right| \right) + \frac{3\,{\gamma}^2}{r_d}\Bigg[\sec^{-1}\left({r_1\over r_d}\right)\\
&& +\sec^{-1}\left({r_2\over r_d}\right)\Bigg]
 +\frac{2\Lambda}{9}\left(r_1^3+r_2^3\right)\,.
\end{eqnarray}

Note that if we consider the limit ${M} 	\rightarrow M_{\odot}$ and $\Lambda 	\rightarrow  0$, we recover the classical result of GR; that is, $\Delta t_{\text{GR}}=4M_{\odot}\left[ 1+ \ln\left(\frac{4r_1r_2}{r_d^2}\right)\right]$. For a round trip from Earth to Mars and back, we get (for $r_d \ll r_1 , r_2$), where $r_1 \approx r_2=2.25\times 10^{11}$ m. is the average distance Earth-Mars. Considering $r_d$, as closest approach to the Sun, like the radius  of the Sun ($R_{\odot} \approx 
 6.960\times 10^{8}$ m) plus the solar corona ($  \sim 10^{9}$ m), $r_d \approx1.696\times 10^{9}$ m, then,  the time delay is 
 ${\Delta t_{\text{GR}} \over c}\approx 240\; \mu\,s\,.$
 To give an idea of the experimental possibilities, we mention that the error in the time measurement of a circuit during the Viking mission was only about $10\, ns$ \cite{Straumann}. If the Horndeski term contributes, then
 \begin{eqnarray}
     \Delta t_{\text{Horndeski}}=\frac{3\,{\gamma}^2}{r_d}\Bigg[\sec^{-1}\left({r_1\over r_d}\right)+\sec^{-1}\left({r_2\over r_d}\right)\Bigg]\,.
 \end{eqnarray}
 
 For a round trip from Earth to Mars and back, we get (for $r_d \ll r_1 , r_2$)
 \begin{eqnarray}
 {\Delta t_{\text{Horndeski}} \over c} =\frac{6\,{\gamma}^2}{r_d\,c}\sec^{-1}\left({r_1\over r_d}\right),
 \end{eqnarray}
it does so that ${\Delta t_{\text{Horndeski}} \over c} < 10\, ns=10^{-8}\,s$, or	$\abs{\gamma} < 23283$ m.

\section{Lyapunov exponents}
\label{LE}

\begin{figure}[!h]
	\begin{center}
	\includegraphics[width=90mm]{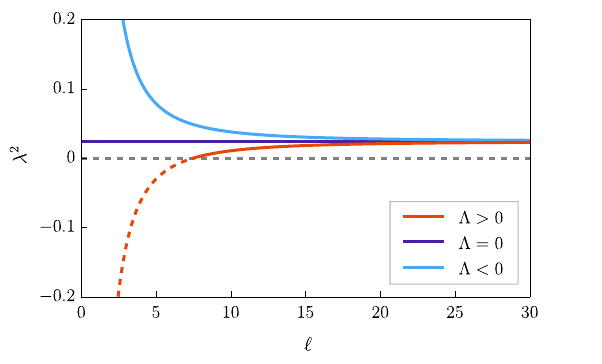}
	\end{center}
	\caption{The Lyapunov exponent $\lambda^2$ as a function of $\ell$. Here,  $M = 1$,  $\gamma=2$, and {\bf{$\ell^2= |3/\Lambda|$.}}}
	\label{Lyapunov}
\end{figure}

Lyapunov exponents are a measure of the average rate at which nearby trajectories converge or diverge in the phase space. Thus, in order to calculate the Lyapunov exponent, we use the Jacobian matrix method \cite{Cardoso:2008bp,Pradhan:2012rkk,Pradhan:2013bli}. So, taking the phase space  $(r,\pi_r)$ the Jacobian matrix $K_{ij}$ is 

\begin{eqnarray}
    K_{11}= \frac{\partial F_1}{\partial r}\,, \quad  K_{12}= \frac{\partial F_1}{\partial \pi_r}\,, \\
    K_{21}= \frac{\partial F_2}{\partial r}\,, \quad K_{22}= \frac{\partial F_2}{\partial \pi_r}\,,
\end{eqnarray}
where $F_1(r,\pi_r)=\frac{dr}{dt}$, and $F_2(r,\pi_r)=\frac{d\pi_r}{dt}$. When the circular motion of particles is considered $\pi_r=0$, and the Jacobian matrix which can be reduce to 

 \begin{equation}
     K_{ij}= \begin{pmatrix}
0 & K_{12}\\
K_{21} & 0 
\end{pmatrix}\, ,
 \end{equation}
being the eigenvalues of the Jacobian matrix the Lyapunov exponent $\lambda$
\begin{equation}
    \lambda=\sqrt{K_{12}K_{21}} \,.
\end{equation}
If $\lambda^2>0$ the circular motion is unstable, if $\lambda^2=0$ the circular motion is marginal, and if $\lambda^2<0$ the circular motion is stable. So, for circular null geodesics the Lyapunov exponent is given by \cite{Cardoso:2008bp}

\begin{eqnarray}
    \lambda&=& \frac{1}{\sqrt{2}}\sqrt{-\frac{r_u^2f(r_u)}{L^2} V_{\text{eff}}^{\prime\prime}(r_u)}\\
\notag & = & \frac{1}{\sqrt{2}}\sqrt{\frac{-f(r_u) \left(r_u \left(r_u f''(r_u)-4 f'(r_u)\right)+6 f(r_u)\right)}{r_u^2}}\,,
\end{eqnarray}
that does not depend on $L$. In Fig. \ref{Lyapunov}, we consider $\ell^2= |3/\Lambda|$, and we plot the Lyapunov exponent as a function of $\ell$. Note that for the spacetimes considered, the exponent is positive, for a fixed value of $M$ and $\gamma$, which indicates a divergence between nearby trajectories, i.e., a high sensitivity to initial conditions. So, once a circular orbit is perturbed, the perturbation will increase exponentially, indicating the presence of chaos. In the figure, the dashed red line indicates that the solution does not represent a black hole solution for the values of $0< \ell < 7.494$. This value can be determine by the expression 
\begin{equation}
\ell=\frac{\sqrt{8 \gamma ^4+12 \gamma ^2 M^2 \left(\omega+6\right)+27 M^4 \left(\omega+3\right)}}{\sqrt{2 \gamma ^2+M^2 \left(\omega+3\right)}}\,,
\end{equation}
where $\omega= \sqrt{\frac{8 \gamma ^2}{M^2}+9}$\,. Moreover, we observe that there exists a range of values for $ \Lambda $ such that  
$\lambda_{AdS}^2 > \lambda_{flat}^2 > \lambda_{dS}^2$, and considering the relation between the corresponding impact parameters $
b_u^{\ell} < b_u^0 < b_u^{\Lambda}$,
this suggests that a smaller impact parameter leads to a larger Lyapunov exponent, indicating a stronger chaotic behavior. This trend is consistent with the corresponding energy levels reported: $E_u^{\ell} > E_u^0 > E_u^{\Lambda}$. On the other hand, for larger positive values of $ \Lambda $, the Lyapunov exponent asymptotically approaches the value obtained in the flat case given by
\begin{equation}    
\lambda_{flat}^2=
\frac{32 \left(8 \gamma ^4+\gamma ^2 M^2 \left(7 \omega+33\right)+9 M^4 \left(\omega+3\right)\right)}{M^6 \left(\omega+3\right)^6}\,.
\end{equation}

It is worth mentioning that a general upper bound of chaos in quantum systems has been proposed by Maldacena, Shenker and Standford from quantum field theory, which indicates that the Lyapunov exponent $\lambda$, describing the strength of chaos, has a temperature-dependent upper bound \cite{Maldacena:2015waa}

\begin{equation}\label{bound}
    \lambda \leq \frac{2\pi T}{\hbar}.
\end{equation}

This temperature-dependent ansatz was examined by shock wave gedanken experiments \cite{Shenker:2013pqa,Shenker:2013yza} and the AdS/CFT correspondence \cite{Maldacena:1997re}. With the black hole thermodynamic relationship $\kappa = 2\pi T$ and setting $\hbar = 1$, the chaos bound Eq. (\ref{bound}) has an equivalent form at the horizon

\begin{equation}
    \lambda\leq \kappa,
\end{equation}
where $\kappa$ is the surface gravity of black holes.




\section{Concluding comments}
\label{conclution}

We considered four-dimensional asymptotically flat, dS and AdS Horndeski black holes and analyzed the motion of massless particles in these background with the objective to study the geodesics, and we discuss the effect of the cosmological constant on the motion of particles. 

For the spacetimes considered we have obtained the horizons analytically. The asymptotically flat spacetime is characterized by two roots, one of them corresponds to the event horizon and the other one is negative, the dS spacetime is characterized by four roots, the event horizon, the cosmological horizon, and two negative roots, and the AdS spacetime is characterized by four roots, one of them corresponds to the event horizon, the other one is negative, and two complex conjugate roots. 

Concerning the motion with $L=0$, we have found that the affine parameter $\lambda$
depends on the energy of the massless particle and does not depend on the metric. However, for coordinate time, the test particles always require an infinite
coordinate time to arrive at the event horizon. However, for $\Lambda=0$, the test particles require an infinite
coordinate time to arrive at infinity.  On the other hand, when the spacetime is dS, the test particles require an infinity coordinate time to reach the cosmological horizon $R_{++}$, and when the spacetime is AdS, the test
particles require a finite coordinate time to arrive
at infinity.  So, the effect of the cosmological constant is to introduce a distance limit for $\Lambda>0$, and a time limit for $\Lambda<0$.

Concerning angular null geodesics, we first performed a qualitative analysis of the effective potential to identify the different kinds of orbits allowed, and we showed that the radial acceleration does not depend on the cosmological constant with a maximum in the inflection point of the effective potential.  One of the allowed orbits is the unstable circular orbit, which depends on $M$ and $\gamma$, and does not depend on the cosmological constant. However, the energy of the particle to reach this orbit is met by $E_{u}^{\ell}>E_{u}^0>E_{u}^{\Lambda}$. Here, the analysis of the Lyapunov exponent shows that the exponent is positive for the spacetimes considered, which indicates a divergence between nearby trajectories and the presence of chaos. Other kind of trajectories are the first kind where we study the bending of the light, and we have found exact solutions for the trajectories, the return points, and the deflection distance. Also, we have found the deflection angle $\hat{\alpha}$, where for the spacetimes considered $\hat{\alpha}\rightarrow\infty$, when $r_d\rightarrow r_u$. Also, $\hat{\alpha}\rightarrow 0$, when $E\rightarrow 0$  ($\Lambda=0$), $\hat{\alpha}\rightarrow \hat{\alpha}_{\Lambda}(0) $, when $E\rightarrow 0$ ($\Lambda>0$) and
$\hat{\alpha}\rightarrow 0$, when $E\rightarrow E_{\ell}$ ($\Lambda<0$). The deflection angle is greater when the black hole has a positive cosmological
constant.  

Also, we have completed the geodesics structure for massless particles by considering the relativistic orbits, such as the second-kind trajectories, the critical trajectories, and the geodesics of the capture zone. Mainly, we have found a new kind of second kind trajectories given by the lima\c con with loop, for AdS spacetimes. 

Additionally, we have considered three classical tests of gravity in the solar system in order to  constraint the $\gamma$ parameter. Thus, for the bending of light 
$\gamma = 34541$ (m) (prograde), and $\gamma = 32899$ (m) (retrograde),
for the gravitational redshift $\abs\gamma \leq 0.945$ (m), and
for the Shapiro time delay $\abs{\gamma} < 23283$ (m).

It could be interesting to incorporate constraints from gravitational wave observations or pulsar timing which require a separate and comprehensive analysis, including waveform modeling and timing residual calculations, and we hope to address it in a forthcoming work.

\acknowledgments

We thank the referee for his/her constructive comments as well as for useful comments and suggestions. This work is partially supported by ANID Chile through FONDECYT Grant Nº 1220871  (P.A.G., and Y. V.).


\begin{thebibliography}{999}

 \bibitem{Abbott:2016blz}
 LIGO Scientific and Virgo Collaborations collaboration, B.~P. Abbott
  et~al., Observation of Gravitational Waves from a Binary Black Hole
  Merger,  Phys. Rev. Lett. 116 (2016) 061102.


\bibitem{Abbott:2016nmj}
 VGW151226: Observation of Gravitational Waves from a 22-Solar-Mass
  Binary Black Hole Coalescence,
  Phys. Rev. Lett. 116 (2016) 241103.


\bibitem{Abbott:2017vtc}
{\scshape VIRGO, LIGO Scientific} collaboration, B.~P. Abbott et~al.,
  GW170104: Observation of a 50-Solar-Mass Binary Black Hole Coalescence
  at Redshift 0.2,
  Phys. Rev. Lett.  118 (2017) 221101.


\bibitem{Abbott:2017oio}
 Virgo, LIGO Scientific collaboration, B.~P. Abbott et~al.,
  GW170814: A Three-Detector Observation of Gravitational Waves from a
  Binary Black Hole Coalescence,
  Phys. Rev. Lett. 119 (2017) 141101.

\bibitem{TheLIGOScientific:2017qsa}
 Virgo, LIGO Scientific collaboration, B.~P. Abbott et~al.,
  GW170817: Observation of Gravitational Waves from a Binary Neutron
  Star Inspiral,
  Phys. Rev. Lett. 119 (2017) 161101.

  \bibitem{Fujii:2003pa}
Y.~Fujii, K.~Maeda, {The scalar-tensor theory of gravitation} (Cambridge
  University Press, 2007).

\bibitem{Horndeski:1974wa}
G.W. Horndeski, {Second-order scalar-tensor field equations in a
  four-dimensional space}. Int. J. Theor. Phys. \textbf{10}, 363 (1974).

\bibitem{Ostrogradsky:1850fid}
M.~Ostrogradsky, {M\'{e}moires sur les \'{e}quations diff\'{e}rentielles, relatives au
  probl\`{e}me des isop\'{e}rim\`{e}tres}. Mem. Acad. St. Petersbourg \textbf{6}(4), 385
  (1850).

  \bibitem{Nicolis:2008in}
A.~Nicolis, R.~Rattazzi, E.~Trincherini, {The Galileon as a local modification
  of gravity}. Phys. Rev. D \textbf{79}, 064036 (2009).

\bibitem{Deffayet:2009wt}
C.~Deffayet, G.~Esposito-Farese, A.~Vikman, {Covariant Galileon}. Phys. Rev. D
  \textbf{79}, 084003 (2009).




  \bibitem{Kolyvaris:2011fk}
T.~Kolyvaris, G.~Koutsoumbas, E.~Papantonopoulos, G.~Siopsis, ``Scalar Hair from
  a Derivative Coupling of a Scalar Field to the Einstein Tensor,'' Class.
  Quant. Grav. \textbf{29}, 205011 (2012).

\bibitem{Rinaldi:2012vy}
  M.~Rinaldi,
  ``Black holes with non-minimal derivative coupling,''
  Phys.\ Rev.\ D {\bf 86}, 084048 (2012)
  [arXiv:1208.0103 [gr-qc]].

\bibitem{Kolyvaris:2013zfa}
T.~Kolyvaris, G.~Koutsoumbas, E.~Papantonopoulos, G.~Siopsis, ``Phase Transition
  to a Hairy Black Hole in Asymptotically Flat Spacetime'', JHEP \textbf{11},
  133 (2013)

\bibitem{Babichev:2013cya}
E.~Babichev, C.~Charmousis, ``Dressing a black hole with a time-dependent
  Galileon'', JHEP \textbf{08}, 106 (2014)

\bibitem{Charmousis:2014zaa}
C.~Charmousis, T.~Kolyvaris, E.~Papantonopoulos, M.~Tsoukalas, ``Black Holes in
  Bi-scalar Extensions of Horndeski Theories'', JHEP \textbf{07}, 085 (2014)



\bibitem{Babichev:2017guv}
E.~Babichev, C.~Charmousis and A.~Leh\'ebel,
``Asymptotically flat black holes in Horndeski theory and beyond,''
JCAP \textbf{04} (2017), 027
[arXiv:1702.01938 [gr-qc]].


\bibitem{Bergliaffa:2021diw}
S.~E.~P.~Bergliaffa, R.~Maier and N.~d.~Silvano,
[arXiv:2107.07839 [gr-qc]].


\bibitem{Walia:2021emv}
R.~K.~Walia, S.~D.~Maharaj and S.~G.~Ghosh,
Eur. Phys. J. C \textbf{82}, 547 (2022)
doi:10.1140/epjc/s10052-022-10451-5
[arXiv:2109.08055 [gr-qc]].


\bibitem{Jha:2022tdl}
S.~K.~Jha, M.~Khodadi, A.~Rahaman and A.~Sheykhi,
Phys. Rev. D \textbf{107}, no.8, 084052 (2023)
doi:10.1103/PhysRevD.107.084052
[arXiv:2212.13051 [gr-qc]].




\bibitem{Amendola:1993uh}
  L.~Amendola,
  ``Cosmology with nonminimal derivative couplings,''
  Phys.\ Lett.\  B {\bf 301}, 175 (1993)
  [arXiv:gr-qc/9302010].




\bibitem{Sushkov:2009hk}
  S.~V.~Sushkov,
  ``Exact cosmological solutions with nonminimal derivative coupling,''
  Phys.\ Rev.\  D {\bf 80}, 103505 (2009)
  [arXiv:0910.0980 [gr-qc]].

\bibitem{Germani:2010hd}
C.~Germani, A.~Kehagias, {UV-Protected Inflation}. Phys. Rev. Lett.
  \textbf{106}, 161302 (2011)

\bibitem{Saridakis:2010mf}
E.N. Saridakis, S.V. Sushkov, {Quintessence and phantom cosmology with
  non-minimal derivative coupling}. Phys. Rev. D \textbf{81}, 083510 (2010)

\bibitem{Huang:2014awa}
Y.~Huang, Q.~Gao, Y.~Gong, {The Phase-space analysis of scalar fields with
  non-minimally derivative coupling}. Eur. J. Phys. C \textbf{75}, 143 (2015)

\bibitem{Yang:2015pga}
N.~Yang, Q.~Fei, Q.~Gao, Y.~Gong, {Inflationary models with non-minimally
  derivative coupling}. Class. Quant. Grav. \textbf{33}(20), 205001 (2016)

\bibitem{Koutsoumbas:2013boa}
G.~Koutsoumbas, K.~Ntrekis, E.~Papantonopoulos, {Gravitational Particle
  Production in Gravity Theories with Non-minimal Derivative Couplings}. JCAP
  \textbf{08}, 027 (2013)


\bibitem{Babichev:2025ric}
E.~Babichev, G.~Esposito-Far{\`e}se, I.~Sawicki and L.~G.~Trombetta,
``Large black-hole scalar charges induced by cosmology in Horndeski theories,''
Phys. Rev. D \textbf{112} (2025) no.2, 024043
[arXiv:2504.07882 [gr-qc]].



\bibitem{Myung:2025wmw}
Y.~S.~Myung,
``Extended thermodynamic analysis of a charged Horndeski black hole,''
Phys. Lett. B \textbf{866} (2025), 139523
[arXiv:2503.01051 [gr-qc]].



  \bibitem{Germani:2010gm}
C.~Germani, A.~Kehagias, {New Model of Inflation with Non-minimal Derivative
  Coupling of Standard Model Higgs Boson to Gravity}. Phys. Rev. Lett.
  \textbf{105}, 011302 (2010)

\bibitem{Germani:2011ua}
C.~Germani, Y.~Watanabe, {UV-protected (Natural) Inflation: Primordial
  Fluctuations and non-Gaussian Features}. JCAP \textbf{1107}, 031 (2011).
\newblock [Addendum: JCAP1107,A01(2011)]

\bibitem{Lombriser:2015sxa}
L.~Lombriser, A.~Taylor, {Breaking a Dark Degeneracy with Gravitational Waves}.
  JCAP \textbf{1603}(03), 031 (2016)

\bibitem{Lombriser:2016yzn}
L.~Lombriser, N.A. Lima, {Challenges to Self-Acceleration in Modified Gravity
  from Gravitational Waves and Large-Scale Structure}. Phys. Lett. B
  \textbf{765}, 382 (2017)

\bibitem{Bettoni:2016mij}
D.~Bettoni, J.M. Ezquiaga, K.~Hinterbichler, M.~Zumalac\'{a}rregui, {Speed of
  Gravitational Waves and the Fate of Scalar-Tensor Gravity}. Phys. Rev. D
  \textbf{95}(8), 084029 (2017)

\bibitem{Baker:2017hug}
  T.~Baker, E.~Bellini, P.~G.~Ferreira, M.~Lagos, J.~Noller and I.~Sawicki,
 ``Strong constraints on cosmological gravity from GW170817 and GRB 170817A,''
  Phys.\ Rev.\ Lett.\  {\bf 119}, no. 25, 251301 (2017)
  [arXiv:1710.06394 [astro-ph.CO]].

\bibitem{Creminelli:2017sry}
  P.~Creminelli and F.~Vernizzi,
 ``Dark Energy after GW170817 and GRB170817A,''
  Phys.\ Rev.\ Lett.\  {\bf 119}, no. 25, 251302 (2017)
  [arXiv:1710.05877 [astro-ph.CO]].

\bibitem{Sakstein:2017xjx}
  J.~Sakstein and B.~Jain,
 ``Implications of the Neutron Star Merger GW170817 for Cosmological Scalar-Tensor Theories,''
  Phys.\ Rev.\ Lett.\  {\bf 119}, no. 25, 251303 (2017)
  [arXiv:1710.05893 [astro-ph.CO]].

\bibitem{Ezquiaga:2017ekz}
  J.~M.~Ezquiaga and M.~Zumalacárregui,
``Dark Energy After GW170817: Dead Ends and the Road Ahead,''
  Phys.\ Rev.\ Lett.\  {\bf 119}, no. 25, 251304 (2017)
  [arXiv:1710.05901 [astro-ph.CO]].

\bibitem{Monitor:2017mdv}
B.~P.~Abbott \textit{et al.} [LIGO Scientific, Virgo, Fermi-GBM and INTEGRAL],
``Gravitational Waves and Gamma-rays from a Binary Neutron Star Merger: GW170817 and GRB 170817A,''
Astrophys. J. Lett. \textbf{848}, no.2, L13 (2017)
[arXiv:1710.05834 [astro-ph.HE]].

\bibitem{Deffayet:2009mn}
C.~Deffayet, S.~Deser and G.~Esposito-Farese,
``Generalized Galileons: All scalar models whose curved background extensions maintain second-order field equations and stress-tensors,''
Phys. Rev. D \textbf{80}, 064015 (2009)
[arXiv:0906.1967 [gr-qc]].

\bibitem{Gong:2017kim}
  Y.~Gong, E.~Papantonopoulos and Z.~Yi,
  ``Constraints on scalar–tensor theory of gravity by the recent observational results on gravitational waves,''
  Eur.\ Phys.\ J.\ C {\bf 78}, no. 9, 738 (2018)
  [arXiv:1711.04102 [gr-qc]].

  \bibitem{Chakraborty:2012sd}
S.~Chakraborty, S.~SenGupta, ``Solar system constraints on alternative gravity
  theories''. Phys. Rev. D \textbf{89}(2), 026003 (2014)

\bibitem{Cruz:2004ts}
N.~Cruz, M.~Olivares and J.~R.~Villanueva,
``The Geodesic structure of the Schwarzschild anti-de Sitter black hole,''
Class. Quant. Grav. \textbf{22} (2005), 1167-1190
[arXiv:gr-qc/0408016 [gr-qc]].



\bibitem{Vasudevan:2005js}
  M.~Vasudevan and K.~A.~Stevens,
  ``Integrability of particle motion and scalar field propagation in Kerr-(Anti) de Sitter black hole spacetimes in all dimensions,''
  Phys.\ Rev.\ D {\bf 72} (2005) 124008
  [gr-qc/0507096].

\bibitem{Hackmann:2008zz}
  E.~Hackmann and C.~Lammerzahl,
  ``Geodesic equation in Schwarzschild- (anti-) de Sitter space-times: Analytical solutions and applications,''
  Phys.\ Rev.\ D {\bf 78} (2008) 024035
  [arXiv:1505.07973 [gr-qc]].


\bibitem{Hackmann:2008zza}
  E.~Hackmann and C.~Lammerzahl,
  ``Complete Analytic Solution of the Geodesic Equation in Schwarzschild- (Anti-) de Sitter Spacetimes,''
  Phys.\ Rev.\ Lett.\  {\bf 100} (2008) 171101
  [arXiv:1505.07955 [gr-qc]].



\bibitem{Olivares:2011xb}
  M.~Olivares, J.~Saavedra, J.~R.~Villanueva and C.~Leiva,
  ``Motion of charged particles on the Reissner-Nordstr\'om (Anti)-de Sitter black holes,''
  Mod.\ Phys.\ Lett.\ A {\bf 26} (2011) 2923
  [arXiv:1101.0748 [gr-qc]].

\bibitem{Cruz:2011yr}
  N.~Cruz, M.~Olivares, J.~Saavedra and J.~R.~Villanueva,
  ``Null geodesics in the Reissner-Nordstrom Anti-de Sitter black holes,''
  arXiv:1111.0924 [gr-qc].

\bibitem{Larranaga:2011fp}
  A.~Larranaga,
 ``Geodesic Structure of the Noncommutative Schwarzschild Anti-de Sitter Black Hole I: Timelike Geodesics,''
  Rom.\ J.\ Phys.\  {\bf 58} (2013) 50
  [arXiv:1110.0778 [gr-qc]].

\bibitem{Villanueva:2013zta}
  J.~R.~Villanueva, J.~Saavedra, M.~Olivares and N.~Cruz,
 ``Photons motion in charged Anti-de Sitter black holes,''
  Astrophys.\ Space Sci.\  {\bf 344} (2013) 437.





\bibitem{Gonzalez:2013aca}
  P.~A.~Gonzalez, E.~Papantonopoulos, J.~Saavedra and Y.~Vasquez,
 `Four-Dimensional Asymptotically AdS Black Holes with Scalar Hair,''
  JHEP {\bf 1312} (2013) 021
  [arXiv:1309.2161 [gr-qc]].





\bibitem{Gonzalez:2015jna}
  P.~A.~Gonzalez, M.~Olivares and Y.~Vasquez,
  ``Motion of particles on a Four-Dimensional Asymptotically AdS Black Hole with Scalar Hair,''
  Eur.\ Phys.\ J.\ C {\bf 75}, no. 10, 464 (2015)
  [arXiv:1507.03610 [gr-qc]].


\bibitem{Ahmed:2025sav}
F.~Ahmed, A.~Al-Badawi and I.~Sakall{\i},
``AdS black strings in a cosmic web: geodesics, shadows, and thermodynamics,''
Eur. Phys. J. C \textbf{85}, no.5, 554 (2025)
[arXiv:2505.13833 [gr-qc]].


\bibitem{Ahmed:2025evv}
F.~Ahmed, A.~Al-Badawi and {\.I}.~Sakall{\i},
``Probing quantum gravity effects: Geodesic structure and thermodynamics of deformed Schwarzschild AdS black holes surrounded by cosmic strings,''
Phys. Dark Univ. \textbf{48} (2025), 101925


\bibitem{Ahmed:2025iqz}
F.~Ahmed, A.~Al-Badawi and {\.I}.~Sakall{\i},
``Exploring geodesics, quantum fields and thermodynamics of Schwarzschild-AdS black hole with a global monopole in non-commutative geometry,''
Nucl. Phys. B \textbf{1017} (2025), 116951
doi:10.1016/j.nuclphysb.2025.116951





  \bibitem{Bhattacharya:2016naa}
S.~Bhattacharya, S.~Chakraborty, {Constraining some Horndeski gravity
  theories}. Phys. Rev. D \textbf{95}(4), 044037 (2017)

\bibitem{Gonzalez:2020vzl}
P.~A.~Gonz\'alez, M.~Olivares, E.~Papantonopoulos and Y.~V\'asquez,
``Constraints on scalar\textendash{}tensor theory of gravity by solar system tests,''
Eur. Phys. J. C \textbf{80} (2020) no.10, 981
[arXiv:2002.03394 [gr-qc]].





\bibitem{Battista:2022krl}
E.~Battista and G.~Esposito,
``Geodesic motion in Euclidean Schwarzschild geometry,''
Eur. Phys. J. C \textbf{82}, no.12, 1088 (2022).
[arXiv:2202.03763 [gr-qc]].








\bibitem{Kuniyal:2015uta}
R.~S.~Kuniyal, R.~Uniyal, H.~Nandan and K.~D.~Purohit,
``Null Geodesics in a Magnetically Charged Stringy Black Hole Spacetime,''
Gen. Rel. Grav. \textbf{48}, no.4, 46 (2016)
[arXiv:1509.05131 [gr-qc]].

\bibitem{Soroushfar:2016yea}
S.~Soroushfar, R.~Saffari and E.~Sahami,
``Geodesic equations in the static and rotating dilaton black holes: Analytical solutions and applications,''
Phys. Rev. D \textbf{94}, no.2, 024010 (2016)
[arXiv:1601.03143 [gr-qc]].

\bibitem{Garfinkle:1990qj}
D.~Garfinkle, G.~T.~Horowitz and A.~Strominger,
``Charged black holes in string theory,''
Phys. Rev. D \textbf{43}, 3140 (1991)
[erratum: Phys. Rev. D \textbf{45}, 3888 (1992)]

\bibitem{Amaro:2020xro}
D.~Amaro and A.~Mac\'\i{}as,
``Geodesic structure of the Euler-Heisenberg static black hole,''
Phys. Rev. D \textbf{102}, no.10, 104054 (2020)

\bibitem{Chen:2022tbb}
D.~Chen and C.~Gao,
``Angular momentum and chaos bound of charged particles around Einstein\textendash{}Euler\textendash{}Heisenberg AdS black holes,''
New J. Phys. \textbf{24}, no.12, 123014 (2022)
[arXiv:2205.08337 [hep-th]].

\bibitem{Cardoso:2008bp}
V.~Cardoso, A.~S.~Miranda, E.~Berti, H.~Witek and V.~T.~Zanchin,
``Geodesic stability, Lyapunov exponents and quasinormal modes,''
Phys. Rev. D \textbf{79} (2009) no.6, 064016
[arXiv:0812.1806 [hep-th]].

\bibitem{Myers:1986un}
R.~C.~Myers and M.~J.~Perry,
``Black Holes in Higher Dimensional Space-Times,''
Annals Phys. \textbf{172}, 304 (1986)

\bibitem{Pretorius:2007jn}
F.~Pretorius and D.~Khurana,
``Black hole mergers and unstable circular orbits,''
Class. Quant. Grav. \textbf{24}, S83-S108 (2007)
[arXiv:gr-qc/0702084 [gr-qc]].

\bibitem{Sperhake:2008ga}
U.~Sperhake, V.~Cardoso, F.~Pretorius, E.~Berti and J.~A.~Gonzalez,
``The High-energy collision of two black holes,''
Phys. Rev. Lett. \textbf{101}, 161101 (2008)
[arXiv:0806.1738 [gr-qc]].

\bibitem{Shibata:2008rq}
M.~Shibata, H.~Okawa and T.~Yamamoto,
``High-velocity collision of two black holes,''
Phys. Rev. D \textbf{78}, 101501 (2008)
[arXiv:0810.4735 [gr-qc]].


\bibitem{Zhen:2025nah}
W.~Q.~Zhen, H.~Guo, M.~H.~Wu and X.~M.~Kuang,
``Orbital precession and Lense-Thirring effect of Horndeski rotating spacetimes,''
Phys. Lett. B \textbf{862} (2025), 139307



     \bibitem{Chandrasekhar:579245}
S. Chandrasekhar,
``The mathematical theory of black holes", Oxford University Press, 2002.


  
  
  
  
  
  

 
\bibitem{Villanueva:2018kem}
  J.~R.~Villanueva, F.~Tapia, M.~Molina and M.~Olivares,
  ``Null paths on a toroidal topological black hole in conformal Weyl gravity,''
  Eur.\ Phys.\ J.\ C {\bf 78} (2018) no.10,  853
  [arXiv:1808.04298 [gr-qc]].

\bibitem{wald}
Wald R.M.:
General relativity.
The University Chicago Press, Chicago (1984).



\bibitem{Gonzalez:2020zfd}
P.~A.~Gonz\'alez, M.~Olivares, Y.~V\'asquez and J.~R.~Villanueva,
``Null geodesics in five-dimensional Reissner-Nordstr$\ddot{o}$m anti-de Sitter black hole,''
Eur. Phys. J. C \textbf{81} (2021) no.3, 236
[arXiv:2010.01442 [gr-qc]].

\bibitem{Straumann}
N. Straumann,
``General relativity and relativistic astrophysics'',
 Springer, Berlin, 1984.




\bibitem{Kagramanova:2006ax}
  V.~Kagramanova, J.~Kunz and C.~Lammerzahl,
  ``Solar system effects in Schwarzschild-de Sitter spacetime,''
  Phys.\ Lett.\ B {\bf 634}, 465 (2006)
  [gr-qc/0602002].

\bibitem{Roy:2019ijp}
S.~Roy and A.~K.~Sen,
``Study of gravitational deflection of light ray,''
J. Phys. Conf. Ser. \textbf{1330} (2019) no.1, 012002.

\bibitem{Fathi:2025byw}
M.~Fathi and A.~{\"O}vg{\"u}n,
``Black hole with global monopole charge in self-interacting Kalb-Ramond field,''
Eur. Phys. J. Plus \textbf{140} (2025) no.4, 280
[arXiv:2501.09899 [gr-qc]].


\bibitem{Sen:2021wld}
A.~A.~Sen, S.~A.~Adil and S.~Sen,
``Do cosmological observations allow a negative {\ensuremath{\Lambda}}?,''
Mon. Not. Roy. Astron. Soc. \textbf{518} (2022) no.1, 1098-1105
[arXiv:2112.10641 [astro-ph.CO]].




\bibitem{Menci:2024rbq}
N.~Menci, S.~A.~Adil, U.~Mukhopadhyay, A.~A.~Sen and S.~Vagnozzi,
``Negative cosmological constant in the dark energy sector: tests from JWST photometric and spectroscopic observations of high-redshift galaxies,''
JCAP \textbf{07} (2024), 072
[arXiv:2401.12659 [astro-ph.CO]].


  
\bibitem{Vessot:1980zz}
  R.~F.~C.~Vessot {\it et al.},
  ``Test of Relativistic Gravitation with a Space-Borne Hydrogen Maser,''
  Phys.\ Rev.\ Lett.\  {\bf 45}, 2081 (1980).





\bibitem{Pradhan:2012rkk}
P.~Pradhan,
``Stability analysis and quasinormal modes of Reissner\textendash{}Nordstr\o{}m space-time via Lyapunov exponent,''
Pramana \textbf{87} (2016) no.1, 5
[arXiv:1205.5656 [gr-qc]].

\bibitem{Pradhan:2013bli}
P.~P.~Pradhan,
``Lyapunov Exponent and Charged Myers Perry Spacetimes,''
Eur. Phys. J. C \textbf{73} (2013) no.6, 2477
[arXiv:1302.2536 [gr-qc]].

\bibitem{Maldacena:2015waa}
J.~Maldacena, S.~H.~Shenker and D.~Stanford,
``A bound on chaos,''
JHEP \textbf{08} (2016), 106
[arXiv:1503.01409 [hep-th]].

\bibitem{Shenker:2013pqa}
S.~H.~Shenker and D.~Stanford,
``Black holes and the butterfly effect,''
JHEP \textbf{03} (2014), 067
[arXiv:1306.0622 [hep-th]].


\bibitem{Shenker:2013yza}
S.~H.~Shenker and D.~Stanford,
``Multiple Shocks,''
JHEP \textbf{12} (2014), 046
[arXiv:1312.3296 [hep-th]].


\bibitem{Maldacena:1997re}
J.~M.~Maldacena,
``The Large N limit of superconformal field theories and supergravity,''
Adv. Theor. Math. Phys. \textbf{2} (1998), 231-252
[arXiv:hep-th/9711200 [hep-th]].











  
















 


  
 



 









    \end{thebibliography}
\end{document}